\documentclass[fleqn,usenatbib]{mnras}

\usepackage[T1]{fontenc}

\DeclareRobustCommand{\VAN}[3]{#2}
\let\VANthebibliography\thebibliography
\def\thebibliography{\DeclareRobustCommand{\VAN}[3]{##3}\VANthebibliography}

\usepackage{graphicx}	% Including figure files
\usepackage{amsmath}	% Advanced maths commands
\usepackage{amssymb}	% Extra maths symbols

\DeclareRobustCommand{\ion}[2]{%
\relax\ifmmode
\ifx\testbx\f@series
{\mathbf{#1\,\mathsc{#2}}}\else
{\mathrm{#1\,\mathsc{#2}}}\fi
\else\textup{#1\,{\mdseries\textsc{#2}}}%
\fi}
\newcommand{\HI}{\ion{H}{i}}

\usepackage{newtxtext,newtxmath}

\title[Hydra \HI{}{} and UV Morphometrics]{WALLABY Pilot Survey: Hydra Cluster Galaxies UV and \HI{}{} morphometrics}

\author[B.W. Holwerda]{Benne W. Holwerda$^{1}$\thanks{Contact e-mail: \href{mailto:benne.holwerda@louisville.edu}{benne.holwerda@louisville.edu}},
Frank Bigiel$^{2}$,  
Albert Bosma$^{3}$,
Helene M. Courtois$^{4}$,\and 
Nathan Deg$^{5}$, 
Helga D\'enes$^{6}$,
Ahmed Elagali$^{7}$,
Bi-Qing For$^{7,8}$, 
Baerbel Koribalski$^{9,10,8}$, \and
Denis A. Leahy$^{11}$, 
Karen Lee-Waddell$^{7,12}$, 
\'Angel R. L\'opez-S\'anchez$^{13,15,8}$, \and
Se-Heon Oh$^{16}$, 
Tristan N. Reynolds$^{7,8}$, 
Jonghwan Rhee$^{7,8}$,
Kristine Spekkens$^{17}$, \and
Jing Wang$^{18}$, 
Tobias Westmeier$^{7,8}$, 
and O. Ivy Wong$^{7,8,12}$\\
\\
$^{1}$ University of Louisville, Department of Physics and Astronomy, 102 Natural Science Building, 40292 KY Louisville, USA.\\
$^{2}$ Argelander-Institut f\"ur Astronomie, Universit\"at Bonn, Auf dem H\"ugel 71, 53121 Bonn, Germany\\
$^{3}$ Aix Marseille Univ, CNRS, CNES, LAM, Marseille, France\\
$^{4}$ Univ Lyon, Univ Claude Bernard Lyon 1, IUF, IP2I Lyon, F-69622, Villeurbanne, France\\
$^{5}$ Department of Physics, Engineering Physics, and Astronomy, Queen’s University, Kingston, ON, K7L 3N6, Canada\\
$^{6}$ ASTRON, Netherlands Institute for Radio Astronomy, 7991 PD Dwingeloo, The Netherlands\\
$^{7}$ ICRAR, The University of Western Australia, 35 Stirling Highway, Crawley WA 6009, Australia \\
$^{8}$ ARC Centre of Excellence for All Sky Astrophysics in 3 Dimensions (ASTRO 3D), Australia\\
$^{9}$ Australia Telescope National Facility, CSIRO Space \& Astronomy, PO Box 76, Epping, NSW 1710, Australia\\
$^{10}$ School of Science, Western Sydney University, Locked Bag 1797, Penrith, NSW 2751, Australia\\
$^{11}$ Department of Physics and Astronomy,
University of Calgary, 2500 University Dr. NW, Calgary, T2N 1N4, Canada\\
$^{12}$ Australia Telescope National Facility, CSIRO Space \& Astronomy, PO Box 1130, Bentley, WA 6102, Australia\\
$^{13}$ Australian Astronomical Optics, Macquarie University, 105 Delhi Rd, North Ryde, NSW 2113, Australia\\
$^{14}$ Department of Physics and Astronomy, Macquarie University, NSW 2109, Australia\\
$^{15}$ Macquarie University Research Centre for Astronomy, Astrophysics \& Astrophotonics, Sydney, NSW 2109, Australia\\
$^{16}$ Department of Physics and Astronomy, Sejong University, 209 Neungdong-ro, Gwangjin-gu, Seoul, Republic of Korea\\
$^{17}$ Department of Physics and Space Science, Royal Military College of Canada, P.O. Box 17000, Station Forces, Kingston, Ontario, Canada K7K 7B4\\
$^{18}$ Kavli Institute for Astronomy and Astrophysics, Peking University, Beijing 100871, China\\
}

\date{Accepted XXX. Received YYY; in original form June 2022}

\pubyear{2022}

\begin{document}
\label{firstpage}
\pagerange{\pageref{firstpage}--\pageref{lastpage}}
\maketitle

% Abstract of the paper
\begin{abstract}
Galaxy morphology in atomic hydrogen (\HI{}{}) and in the ultra-violet (UV) are closely linked. This has motivated their combined use to quantify morphology over the full \HI{}{} disk for both \HI{}{} and UV imaging. 
We apply galaxy morphometrics: Concentration, Asymmetry, Gini, $M_{20}$ and Multimode-Intensity-Deviation statistics to the first moment-0 maps of the WALLABY survey of galaxies in the Hydra cluster center. Taking advantage of this new \HI{}{} survey, we apply the same morphometrics over the full \HI{}{} extent on archival GALEX \textit{FUV} and \textit{NUV} data to explore how well \HI{}{} truncated, extended ultraviolet disk (XUV) and other morphological phenomena can be captured using pipeline WALLABY data products. 
Extended \HI{}{} and UV disks can be identified relatively straightforward from their respective concentration. Combined with WALLABY \HI{}{}, even the shallowest GALEX data is sufficient to identify XUV disks. 
% We find no correlation with distance from the center of the Hydra cluster and \HI{}{} morphology. 
% we're looking at statmorph GALEX and \HI{}{} morphometrics to explore the quantified relations of \HI{}{} and UV disk appearance with position in the cluster.
Our second goal is to isolate galaxies undergoing ram-pressure stripping in the \HI{} morphometric space. We employ four different machine learning techniques, a decision tree, a k-nearest neighbour, a support-vector machine, and a random forest. Up to 80\% precision and recall are possible with the Random Forest giving the most robust results. 
% Galaxies undergoing ram-pressure stripping and those who are not, are can be isolated in \HI{}{} morphology space, using either a decision tree or a support vector machine. 
\end{abstract}

% Don't make up new ones.

%%%%%%%%%%%%%%%%%%%%%%%%%%%%%%%%%%%%%%%%%%%%%%%%%%

%%%%%%%%%%%%%%%%% BODY OF PAPER %%%%%%%%%%%%%%%%%%

\section{Introduction}

In principle, the appearance of galaxies is a direct result of the processes that formed and shaped them. Turning their appearance into a qualified value, while accounting for viewing angle is a long-standing undertaking in observational extragalactic astronomy. 

While most morphology quantifications focus on the stellar component of galaxies, a very information-rich view of gas-rich galaxies is the 21cm line emission of neutral hydrogen (\HI{}{}). This gas spans most often a larger disk than the stars and it is considered more sensitive to early interaction \citep[e.g.,][]{Hibbard01}. 
There is an active interest in the outskirts of spiral galaxy disks because they are the sites of the most recent, readily observable acquisition of gas for these systems \citep[e.g.][]{Sancisi08}, as well as low-level star-formation \cite[e.g.,][]{Dong08,Koribalski09, Bigiel10b, Alberts11, Lopez-Sanchez15, Watson16}.

This low-level star-formation was first discovered in H$\alpha$ emission by \cite{Ferguson98a} and \cite{Lelievre00} and later in GALEX ultraviolet imaging as Extended UltraViolet (XUV) disks  \citep{Thilker05a,Thilker05b, Thilker07b, Gil-de-Paz05, Gil-de-Paz07, Zaritsky07, Lemonias11,Meurer17,Koribalski17}.  
These XUV disk complexes are generally $\sim100$ Myr old, 
which explains why most lack H$\alpha$ \citep{Alberts11}. UV knots in interactions may be an exception \citep{Lopez-Sanchez15}.
A different explanation, i.e. a top-light Initial Mass Function (IMF), as proposed by \cite{Meurer09,Koda15,Bruzzese15,Watts18}, remains a possibility but the IMF is stochastically sampled. The metallicities of these complexes, as derived from emission lines, are sub-solar, in the range of $0.1-1 ~ Z_\odot$ \cite{Gil-de-Paz07,Bresolin09a, Werk10, Lopez-Sanchez15}.

% Note that in the case of NGC 1512 the ionized gas, H-alpha and the bright emission lines, are detected using spectrscopy in ~70% of the observed UV-rich knots (L-S et al. 2015)

It was noted early by several authors that the atomic hydrogen as observed by the 21cm fine structure line and the ultraviolet structure seem closely related. 
That the \HI{}{} disk extends well beyond the optical disk of spiral galaxies has been known for a long time \cite[e.g.,][]{Bosma78,Begeman89, Meurer96, Meurer98, Swaters02, Noordermeer05, Walter08, Boomsma08, Elson11, Heald11a, Heald11b, Zschaechner11b, de-Blok08} but only recently the close correlation with UV structure drove a direct comparison \citep[cf][]{de-Paz08,Lemonias11}. 

A lopsided appearance of the outer \HI{}{} disk \citep{Jog09,van-Eymeren11a,van-Eymeren11b} or an  asymmetry \citep{Giese16, Reynolds20} can have a myriad of internal or external processes responsible for it: accretion along cosmic web filaments \citep{Bournaud05b}, minor satellite accretion \citep{Zaritsky97}, tidal interactions \citep{Jog09, Koribalski09}, ram-pressure stripping \citep{Moore98} as well as sloshing within the dark matter halo \citep{Stinson09}. The anecdotal similarity and the potential common origin of gas accretion is what drove the use of morphometrics on combined data-sets \citep{Holwerda11b,Holwerda12c}.

% \cite{van-Eymeren11a,van-Eymeren11b}
% \cite{Giese16}
% \cite{Reynolds20} 

Parameterisation of \HI{}{} disk appearance is different from stellar parametrization because the \HI{}{} disk is based on line emission and therefore has a lower dynamic range. The area covered by the disk is larger, but the spatial resolution is typically an order of magnitude lower due to the much larger \HI{} beam (or PSF).

A good companion data-set is typically GALEX ultraviolet (FUV and NUV) because the spatial sampling is often of similar to \HI{} scales, and \HI{} and the UV from newly formed young stars are related in distribution \citep{Thilker05a,Gil-de-Paz05} and possibly linked through mutual formation processes \citep{Heiner14, Lopez-Sanchez18}. The main benefit of using both \HI{}{} and UV imaging information is that UV  images are inherently very clumpy in appearance. It becomes increasingly difficult to distinguish which UV source is a star-forming region belonging to the disk and which is a background source. By using the \HI{}{} outer contour, the extent of the UV disk can be identified more reliably \citep[e.g. the discussion on outermost \ion{H}{ii} regions in][]{Hunter18}.

In this paper we apply the galaxy morphometrics originally developed for stellar disks which were applied with some success on \HI{} data in the past \citep{Holwerda11a,Holwerda11b,Holwerda11c,Holwerda11d,Holwerda11e,Holwerda12c,Giese16,Reynolds20} but on often heterogeneous data. For example \cite{Giese16} pointed out that these depend strongly on the signal-to-noise ratio  (s/n) of each object, complicating their use across surveys or with varying s/n. 
The optimal application is therefore within a single survey although careful comparisons can successfully be made between instrument setups of different surveys \citep{Reynolds20}. We use the {\sc statmorph} implementation of these morphometrics \citep{Rodriguez-Gomez19}. 

Starting with the early WALLABY data \citep{Serra15b,Koribalski20,For19,Kleiner19,Lee-Waddell19,Elagali19,Reynolds19}, we apply the image morphometrics on \HI{} and GALEX FUV data to explore their utility in identifying disturbed \HI{} disks, \HI{} deficient or truncated disks, and extended ultraviolet disks. 
This paper is organized as follows: 
Section \ref{s:wallaby} describes the early WALLABY ASKAP and GALEX data products used,
Section \ref{s:morphometrics} describes the morphometrics employed on both \HI{} and GALEX imaging, 
Section \ref{s:results} describes the results for morphometrics to identify extended and ram-pressure stripped disks,
Section \ref{s:RPS} describes the \HI{} morphometrics of ram-pressure galaxies,
Section \ref{s:ML} describes the machine learning applied to the \HI{} morphometric space to identify RPS galaxies, 
and Section \ref{s:conclusions} briefly discusses all of our results and a future outline of their potential use in WALLABY survey proper.
Throughout we use the Planck (2015) cosmology  \citep[$\mathrm{H_0} = 67.74~  \mathrm{km /s / Mpc},~ \Omega_0 = 0.3075$,][]{Planck-Collaboration15}.

\section{WALLABY Survey}
\label{s:wallaby}

The Widefield ASKAP L-band Legacy All-sky Blind surveY \citep[or WALLABY,][]{Koribalski20} is one of several  key surveys that will soon begin on the Australian SKA Pathfinder \citep[ASKAP][]{ASKAP,askap1,askap2,askap4, McConnell16, Hotan14,Hotan21,Koribalski22}, which is an imaging radio telescope located at the Murchison Radio-astronomy Observatory in Western Australia. The aim of WALLABY is to use the powerful wide-field phased-array technology of ASKAP to observe a large fraction of the whole sky in the 21-cm line of \HI{} at 30-arcsec angular resolution, which will yield information for hundreds of thousands of external galaxies in the local Universe. 

Science goals include an understanding of the role of stellar and black hole feedback, gas accretion and galaxy interactions in galaxy evolution. Currently the WALLABY Pilot Survey observations covered three 60 deg$^2$ fields, of which the Hydra field is one. The Pilot Survey is allowing the team to develop, deploy and commission its pipeline processing, data verification and post-processing algorithms, as well as measure the properties of hundreds of galaxies, including their distance, neutral hydrogen mass, and total mass. 

\begin{figure}
    \centering
    \includegraphics[width=0.5\textwidth]{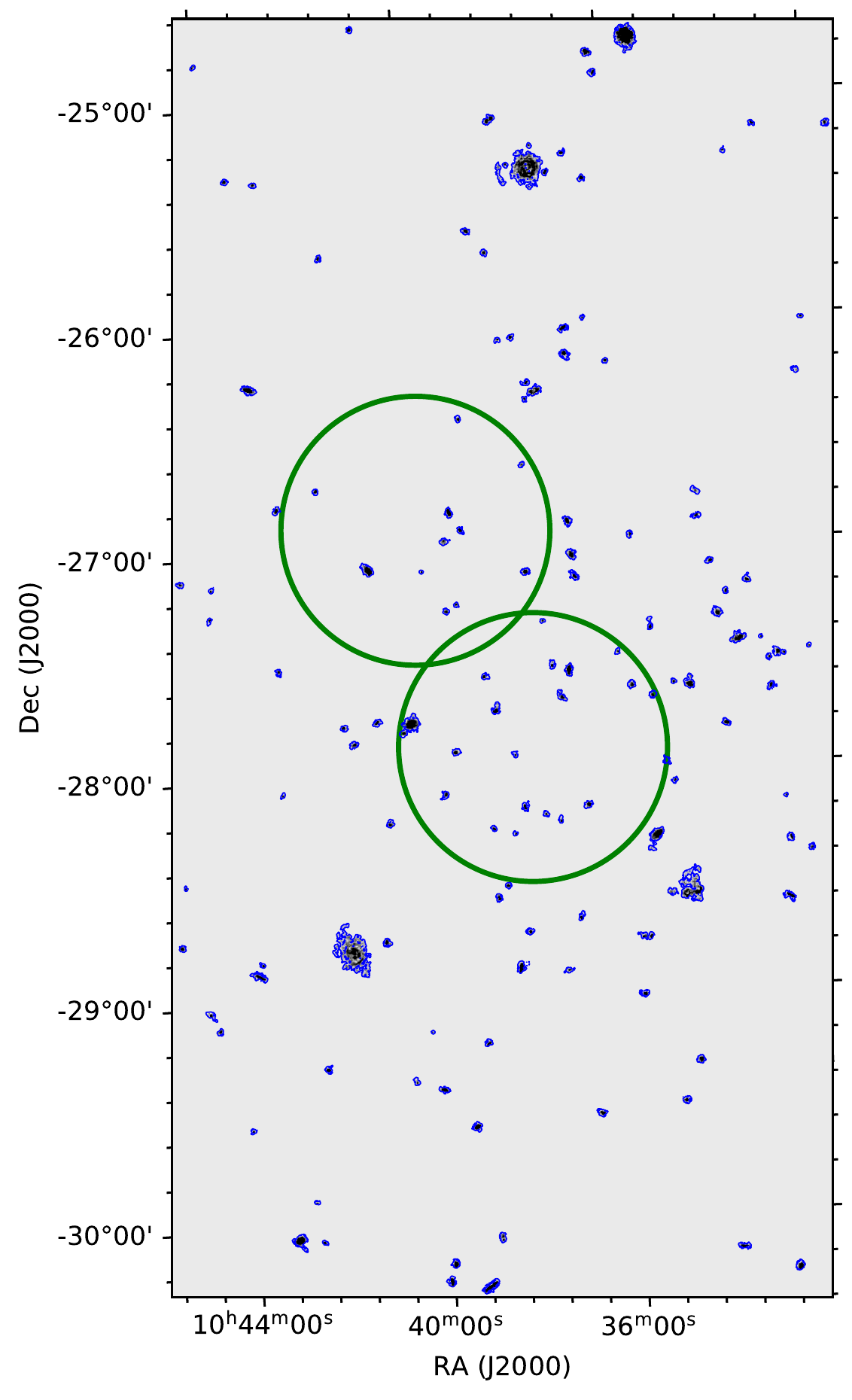}
    \caption{The moment-0 map of the central part of the Hydra cluster with \HI{} detections outlined in blue (set to 10 Jy/beam$\times$Hz in the moment-0). The location of the two GALEX fields are outlined in green. The full TR2 coverage extends to either side.}
    \label{f:mom0}
\end{figure}{}

\subsection{Hydra Cluster Field}

The Hydra field was observed with ASKAP as part of the WALLABY Pilot Survey.
% as an internal data release, including sofia source extraction on the initial data cube. 
The first and second internal data releases (Hydra TR1 and TR2) are available in the sense that the original data cubes produced by the ASKAP pipeline are available to team members on the CSIRO data archive CASDA. In addition, WALLABY has run its own SoFiA software \citep[][]{Serra15a,Westmeier21} to produce source catalogues and source images/spectra for the team to work on.
The WALLABY team internal data release covers a sub-region of the core of the Hydra cluster in a region of the cube that is relatively clean from artefacts \citep{Reynolds21,Wang21}. The region covers the RA range of approximately 10:32:00 to 10:45:00 and the declination range of approximately -30:20:00 to -24:40:00, corresponding to about 5.5 × 3.5 degrees on the sky or roughly one third of the eventual full Hydra field (Figure \ref{f:mom0}). The redshift range considered here covers \mbox{cz = 500} to 15,000 km/s ($z < 0.05$), roughly half of the RFI-free band available to WALLABY.

In order to reduce the impact of residual continuum emission on the source finding, the positions of all NRAO VLA Sky Survey \citep[NVSS][]{Condon98} sources with a flux density in excess of 150 mJy were manually flagged in the form of a box of size $11 \times 11$ pixels centred on the pixel containing each NVSS position.

The great benefit for calculating morphometrics in this field is that the galaxies are all at comparable distances and the observations all probe similar spatial scales.

\subsection{SoFiA Source Finding}

\HI{} detections done with SoFiA \citep{Serra15a,Westmeier21} were linked across a spatial and spectral radius of 2 with a minimum size requirement for a reliable source of 8 spatial pixels and 5 spectral channels. SoFiA’s reliability filter was then applied to remove all detections with a reliability below 0.6, using a Gaussian kernel density estimator of size 0.3 times the covariance. All remaining sources were then parameterised, assuming a restoring beam size of 30 arcsec for all integrated flux measurements.

The resulting detections were inspected by eye to remove obvious artefacts that were unlikely to be genuine \HI{} sources. Many of these artefacts are related to inadequate flagging of the upper-left and lower-left corner beams of the mosaic that have strayed into the region considered here. After manual removal of artefacts, 272 detections remain.
% This is the second data-release (TR2) of the WALLABY collaboration. 
This data set was released internally to the WALLABY team as Hydra TR2. 

%and are herewith released to the WALLABY team as an initial DR1. The source catalogue from the full Hydra field is expected to be released at a later stage once the remaining artefacts have been removed from the image.
% Please note that the remaining 148 \HI{} detections do not necessarily correspond to individual galaxies. On the one hand, there may still be a very small number of artefacts in the catalogue. On the other hand, close galaxy pairs or interacting galaxies with a common \HI{} envelope may be listed as a single catalogue entry. Caution is therefore advised when carrying out scientific analysis based on the presented data.

% \subsection{Data Products used}

We use the internal moment-0 maps (total line intensity across frequencies associated with the galaxy) produced (Figure \ref{f:mom0}), the 3D mask cube, and the SoFiA source catalog. These formed the basis for a segmentation image, an image denoting which pixels belong to an object in the catalog. 
We replaced non-zero pixels in the mask cube, collapsed on the frequency axis, with the source identification in the SoFiA catalog to be used as the segmentation image input in {\sc statmorph} \citep[][see \S \ref{s:morphometrics}]{Rodriguez-Gomez19}. To compute the morphometrics, we use the moment-0 and the collapsed mask cutouts from SoFiA, not the full mosaic out of memory management considerations.

% since use of the full mosaic require an excessive amount of computer memory. 

\begin{figure*}
    \centering
    \includegraphics[width=0.49\textwidth]{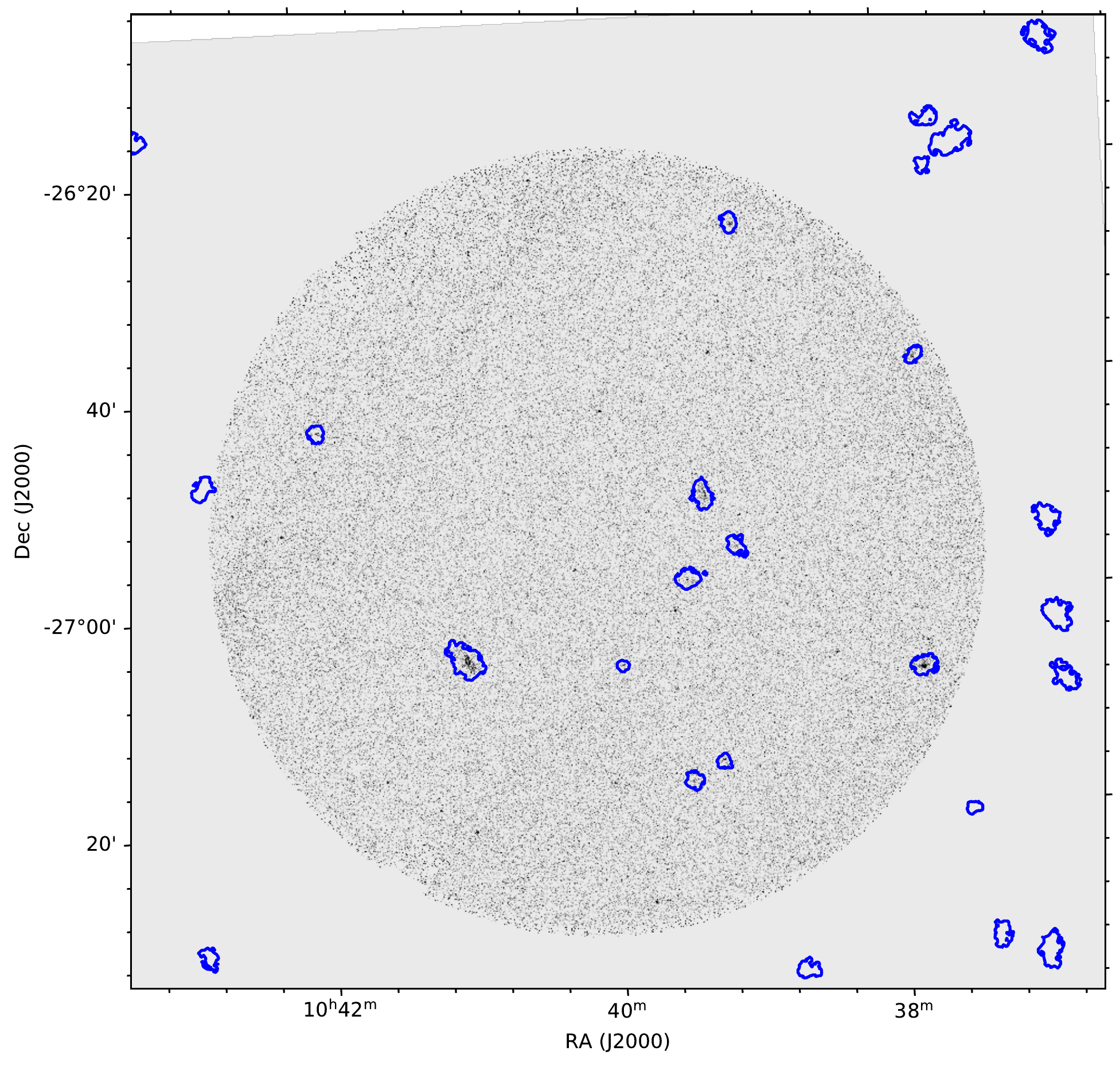}
    \includegraphics[width=0.49\textwidth]{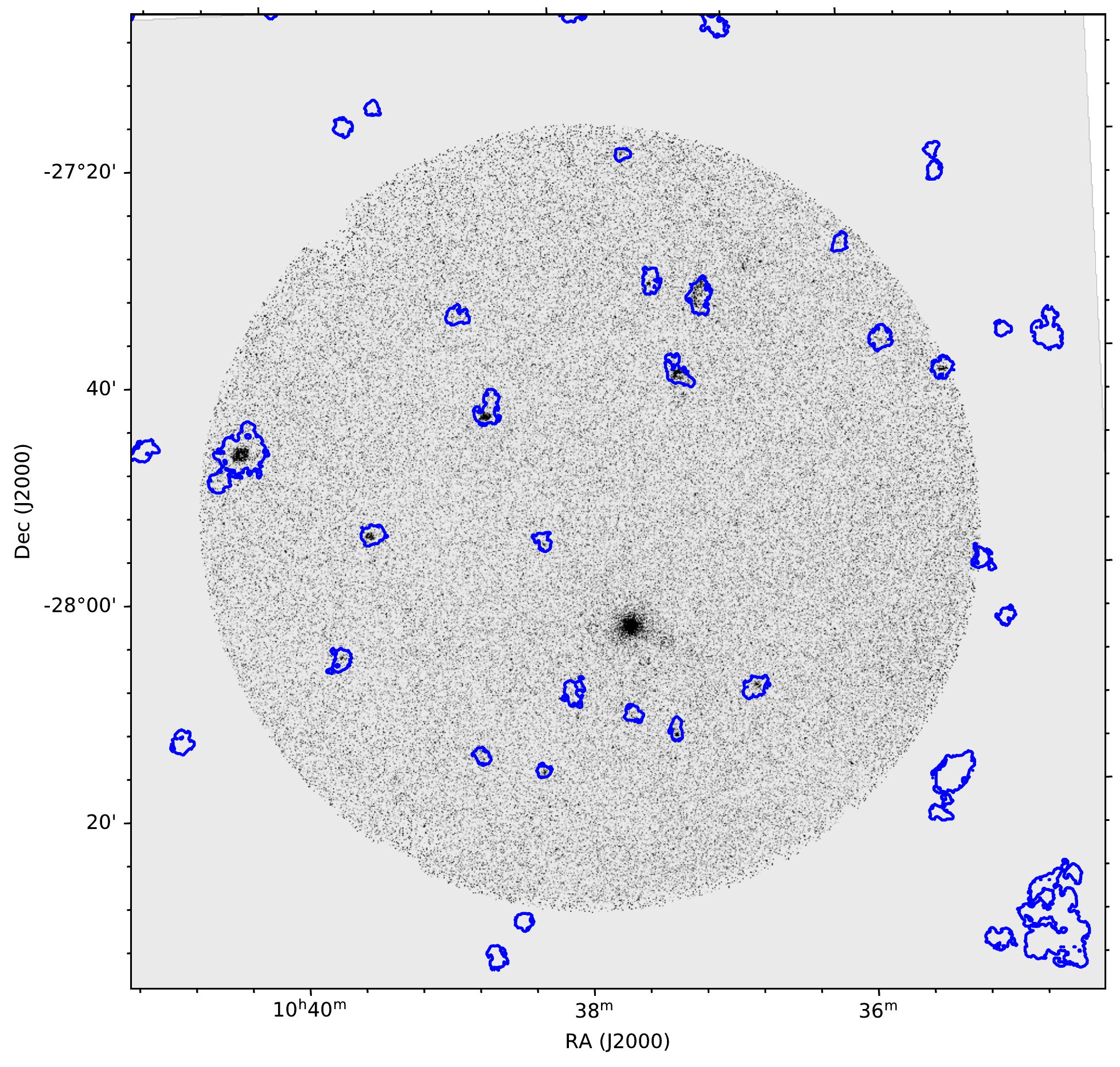}
    \caption{The FUV filter GALEX observations of the Hydra cluster. \HI{} detections are marked in blue (contour at 10 Jy/beam$\times$Hz in the moment-0 map). There is an overlap of 30 \HI{} detections in the GALEX fields with associated GALEX morphometrics. The bright FUV source in the second field is a foreground galactic star.}
    \label{f:galex}
\end{figure*}{}

\subsection{GALEX observations of Hydra}

Two fields were observed for the original All Sky Imaging Surveys (AIS) with the GALEX telescope in both the FUV and NUV channels (1350-1750 and 1750-2800 \AA, respectively) with 210s exposures (Figure \ref{f:galex}). Overlap with the WALLABY central FOV is good but not all the \HI{} detections are covered by UV observations (30 out of 272). While 63 galaxies have non-zero morphometrics, only 30 galaxies overlap with GALEX data, the remainder is the result of morphometrics on the empty sky contribution. Statmorph runs on a segmentation map, in this case the \HI{} one and will measure morphometrics where no source flux is present. Some SWIFT observations exist but only cover a single galaxy in the \HI{} field and thus were not used.

\section{Morphometric Catalogs}
\label{s:morphometrics}

An observational approach to characterizing galaxy appearances is ``non-parametric'' or morphometric parameters. In essence, several authors and groups have tried to come up with parameters that can be applied to images that do not depend too much on resolution or a preconceived idea about the shape of the profile. These non-parametric or ``morphometric'' parameters\footnote{A term first coined here \cite{morphometrics}. The term ``morphometrics'' originates from biology to describe animal appearance in quantitative form, e.g. ``length of fins'' or how far the eyes are set apart.}
can then be used as quantities to classify galaxies along the Hubble Tuning fork or to identify mergers in a population of galaxies. 

The morphometric parameters considered here are Concentration, Asymmetry and Smoothness from \cite{CAS}, $M_{20}$ and Gini from \cite{Lotz04}, and the MID parameters from \cite{Peth16, Rodriguez-Gomez19}. We use the {\sc statmorph} package described in \cite{Rodriguez-Gomez19} to compute the morphometrics. 
We utilize a Gaussian smoothing kernel with a 1 pixel FWHM (6") for both \HI{} and UV implementations of {\sc statmorph}. This choice is not critical for the work here as this input smoothing kernel is only used for the S\'ersic profile fit in {\sc statmorph}, the index of which we do not use here. Both \HI{} and UV profiles are typically not well described with a S\'ersic profile \citep[cf][ and Reynolds et al. in prep. for result of \HI{} radial profiles]{Leroy08,Bigiel11,Wang14,Swaters02}.

% \textbf{maybe better to use 0.5 pixel?}

In many cases, these morphometrics still require inputs other than the image. The relevant input parameters are the central position of the galaxy ($x_c$, $y_c$), and a definition of the area over which these parameters are computed, the segmentation of the image (i.e. which pixels belong to the object of interst). These two measurements, the center of the object and which parts of the image belong to it, are often obtained from  source extractor \citep{SE,seman}. 

Here, the segmentation map was generated from the the SoFiA 3D mask and the central positions were computed by {\sc statmorph}. By using the \HI{} map for the segmentation, the granular nature of the FUV/NUV imaging can be accounted for and the oversegmentation of the UV image can be avoided.
The use of the \HI{} contour effectively isolates distant galaxies from Galactic stellar sources. Foreground Galactic stars are less of a consideration (cf Figure \ref{f:galex}).

\subsection{Concentration-Asymmetry-Smoothness (CAS) Morphometrics}
\label{s:cas}

CAS refers to the now commonly used Concentration-Asymmetry-Smoothness space \citep{CAS} for morphological analysis of distant galaxies. Concentration of the light, symmetry around the centre and smoothness as an indication of substructure.

Concentration is defined by \cite{Bershady00} as:
\begin{equation}
C = 5 ~ \log (r_{80} /  r_{20})
\label{eq:c}
\end{equation}
\noindent with $r_{f}$ as the radius containing percentage $f$ of the light of the galaxy (see definitions of $r_f$ in \cite{SE,seman}).
In the optical, typical values for the concentration index are $C=2-3$ for discs, $C>3.5$ for massive ellipticals, while peculiars span the
entire range \citep{CAS}.

The asymmetry is defined as the level of {\em point}-, (or rotational-) symmetry around the centre of the galaxy \citep{Abraham94,CAS}:
\begin{equation}
A = {\Sigma_{i,j} | I(i,j) - I_{180}(i,j) |  \over \Sigma_{i,j} | I(i,j) |  } - A_{bgr},
\label{eq:a}
\end{equation}
\noindent where $I(i,j)$ is the value of the pixel at the position $[i,j]$ in the image, and $I_{180}(i,j)$ is the pixel at position $[i,j]$ in the galaxy's image, after it was rotated $180^\circ$ around the centre of the galaxy. In the {\sc statmorph} implementation, the asymmetry is calculated in the inner 1.5 Petrosian\footnote{The Petrosian radius is one of several definitions to automatically assign a size and aperture to inherently fuzzy galaxies. For a comprehensive treatment on them, see \cite{Graham05a,Graham05b}.} radii (tyical size of the stellar disk), the background asymmetry is subtracted and A is minimized by moving the center of rotation. This is a different implementation than used in \cite{Holwerda12c}. Note that asymmetry maximum value is 2 (all pixels off-center) and can be negative if the background asymmetry value is large. We note that we do not subtract a background when using the moment-0 \HI{} maps as these are extracted from the field using the 3D source mask.

% Inspired by the ``unsharp masking" technique \citep{Malin78b}, Smoothness is defined by \cite{Takamiya99} and \cite{CAS} as:
% \begin{equation}
% S = {\Sigma_{i,j} | I(i,j) - I_{S}(i,j) | \over \Sigma_{i,j} | I(i,j) | }
% \label{eq:s}
% \end{equation}
% \noindent where $I_{S}(i,j)$ is the same pixel in a smoothed image. What type of smoothing is used has changed over the years. Often a fixed Gaussian smoothing kernel is chosen for simplicity.

The fact that this Smoothness has another input parameter in the form of the size of the smoothing kernel, makes it highly ``tunable'', meaning one gets out exactly what was optimized for. It is very difficult to reliably compare between catalogs or even implementations. For this reason, ``Smoothness'' is not considered further here. 

\subsection{Gini and $M_{20}$}
\label{ss:gm20}

\cite{Abraham03} and \cite{Lotz04} introduce the Gini parameter to quantify the distribution of flux over the pixels in an image.
They use the following definition:
\begin{equation}
G = {1\over \bar{I} n (n-1)} \Sigma_i (2i - n - 1) I_i ,
\label{eq:g}
\end{equation}
\noindent $I_i$ is the value of pixel i in an ordered list of the pixels, $n$ is the number of pixels in the image, and $\bar{I}$ is the mean pixel value in the image. 
This is the computationally least expensive implementation where not the entire mosaic has to be loaded. 

The Gini parameter is an indication of equality in a distribution \citep[initially an economic indicator][]{Gini12,Yitzhaki91}, with G=0 the perfect equality (all pixels have the same fraction of the flux) and G=1 perfect inequality (all the intensity is in a single pixel). Its behaviour is therefore in between that of a structural measure and concentration. Gini appears quite sturdy as it does not require the center of the object to be computed. It remains relatively unchanged, even when the object is lensed \citep{Florian16} and it is popular for this reason. However, it depends strongly on the image's signal-to-noise \citep{Lisker08}. In essence, noise can add pixels with no fraction of the flux in them, artificially increasing the Gini value.

\cite{Lotz04} also introduced a way to parameterize the extent of the light in a galaxy image. They define the spatial second order moment as the product of the intensity with the square of the projected distance to the centre of the galaxy. This gives more weight to emission further out in the disk. It is sensitive to substructures such as spiral arms and star-forming regions but insensitive to whether these are distributed symmetrically or not.
The second order moment of a pixel $i$ is defined as:
\begin{equation}
M_i = I_i \times [(x-x_c)^2 + (y-y_c)^2 ],
\label{eq:Mi}
\end{equation}
where $[x, y]$ is the position of a pixel with intensity value $I_i$ in the image and $[x_c, y_c]$ is the central pixel position of the galaxy in the image. 

The total second order moment of the image is given by:
\begin{equation}
M_{tot} = \Sigma_i M_i = \Sigma I_i [(x_i - x_c)^2 + (y_i - y_c)^2].
\label{eq:mtot}
\end{equation}

\cite{Lotz04} use the relative contribution of the brightest 20\% of the pixels to the second order moment as a measure of disturbance of a galaxy
after sorting the list of pixels by intensity ($I_i$):
\begin{equation}
M_{20} = \ log \left( {\Sigma_i M_i  \over  M_{tot}}\right), ~ {\rm for} ~ \Sigma_i I_i < 0.2 I_{tot}. \\
\label{eq:m20}
\end{equation}
The $M_{20}$ parameter is sensitive to bright regions in the outskirts of disks and higher values can be expected in galaxy images (in the optical and UV) with star-forming outer regions as well as those images of strongly interacting disks. 

\subsection{multimode–intensity–
deviation (MID) morphometrics}
\label{s:mid}

The MID morphometrics \citep{Freeman13,Peth16} were introduced as an alternative to the Gini–M20 and CAS morphometrics to be more sensitive to recent mergers. However, these new morphometrics have not been tested as extensively as the Gini–M20 and CAS statistics, especially using hydrodynamic simulations \citep{Lotz08a, lotz10a,Lotz11b, Bignone17}, see also the discussion in the implementation in {\sc statmorph} \citep{Rodriguez-Gomez19}. In the case of \HI{} data, none of the morphometrics have been extensively tested, and are to be viewed as purely phenomenological. 

We explore them here because the UV morphology tends to be much more ``clumpy'' than the optical one and, by using the \HI{} outer contour, one can still consider such galaxies as a single object. The MID morphometrics were developed to identify mergers by distinct cores, not segregated by the segmentation, and may prove useful.

\begin{figure*}
    \centering
    \includegraphics[width=\textwidth]{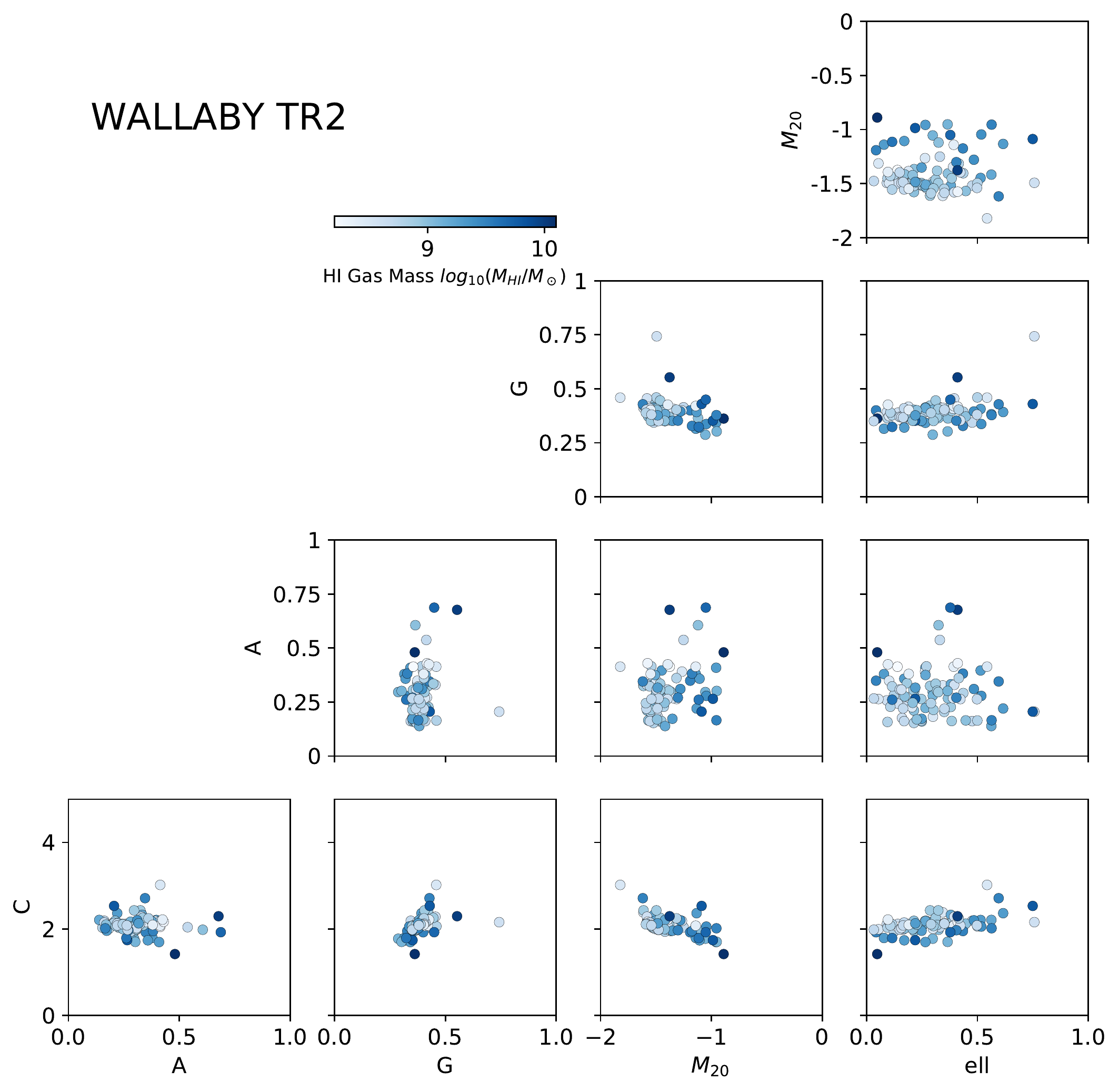}
    \caption{The \HI{} morphometrics corner plot. Concentration, Asymmetry, Gini and M20 together with the ellipticity computed by {\sc statmorph} with the \HI{} mass from WALLABY. }
    \label{f:corner:hi}
\end{figure*}{}

\section{Results}
\label{s:results}

In this section we discuss the two morphometric catalogs computed using the WALLABY \HI{} data and the WALLABY and GALEX data combined. 

\subsection{\HI{} Morphometrics}

Figure \ref{f:corner:hi} shows the distribution of the most commonly used morphometrics for the \HI{} moment-0 map. There is a good range of values, despite the much lower dynamical range typically present in \HI{} maps. The concentration values for \HI{} are clustered around a value of 2. There is a full range of asymmetry values. Asymmetry is calculated with the inclusion of a background component by {\sc statmorph}, but none is included here since the values in the moment-0 map not belonging to an object are set to zero. There are also reasonably high values for the Gini and $M_{20}$ parameters indicating a fair amount of substructure resolved by the WALLABY observations. Concentration and $M_{20}$ are anti-correlated to some degree as can be expected because $M_{20}$ up-weighs pixels far from the center. 
The range of ellipticity values indicates a wide distribution of viewing angles on the \HI{} disks of these galaxies.
These ellipticity values are based on second order moments and not ellipse fitting of the \HI{} cube. The \HI{} morphometrics show a wide range in value. The only parameters that show a narrow range are concentration and Gini, which both can be attributed to the lower dynamic range of the moment-0 maps.

\begin{figure*}
    \centering
    \includegraphics[width=\textwidth]{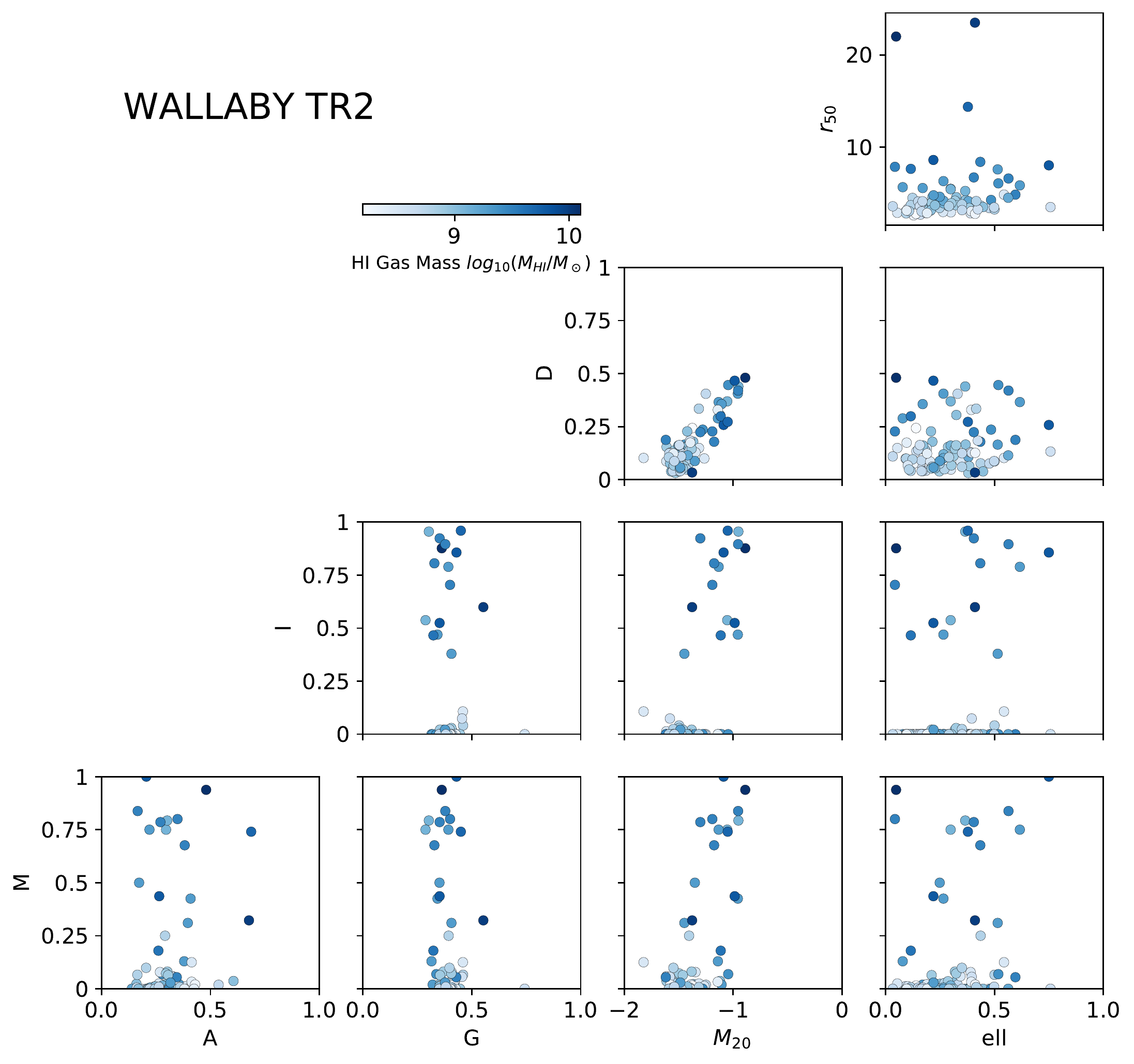}
    \caption{The multimode-intensity-deviation (MID) morphometrics compared against the Concentration-Asymmetry-Gini-$M_{20}$ of the \HI{} morphometrics computed by {\sc statmorph} with the \HI{} mass from WALLABY. }
    \label{f:mid_cross}
\end{figure*}{}

\subsection{MID \HI{} Morphometrics}

The multimode-intensity-deviation (MID) morphometrics for the \HI{} moment-0 maps are compared to the more traditional C-A-G-$M_{20}$ in Figure \ref{f:mid_cross}. Morphometrics were never meant to be an orthogonal parameter space but these MID morphometrics may probe slightly different phenomena. Deviation and $M_{20}$ are clearly correlated and Multimode and Intensity may be weakly correlated with $M_{20}$ as well. Both Multimode and Intensity values are predominantly close to 0 but there are notable outliers. These MID morphometrics, which are sensitive to substructure within an object, may prove useful for \HI{} and UV science in addition to the other, original morphometrics developed for optical morphology.

\begin{figure}
    \centering
    \includegraphics[width=0.5\textwidth]{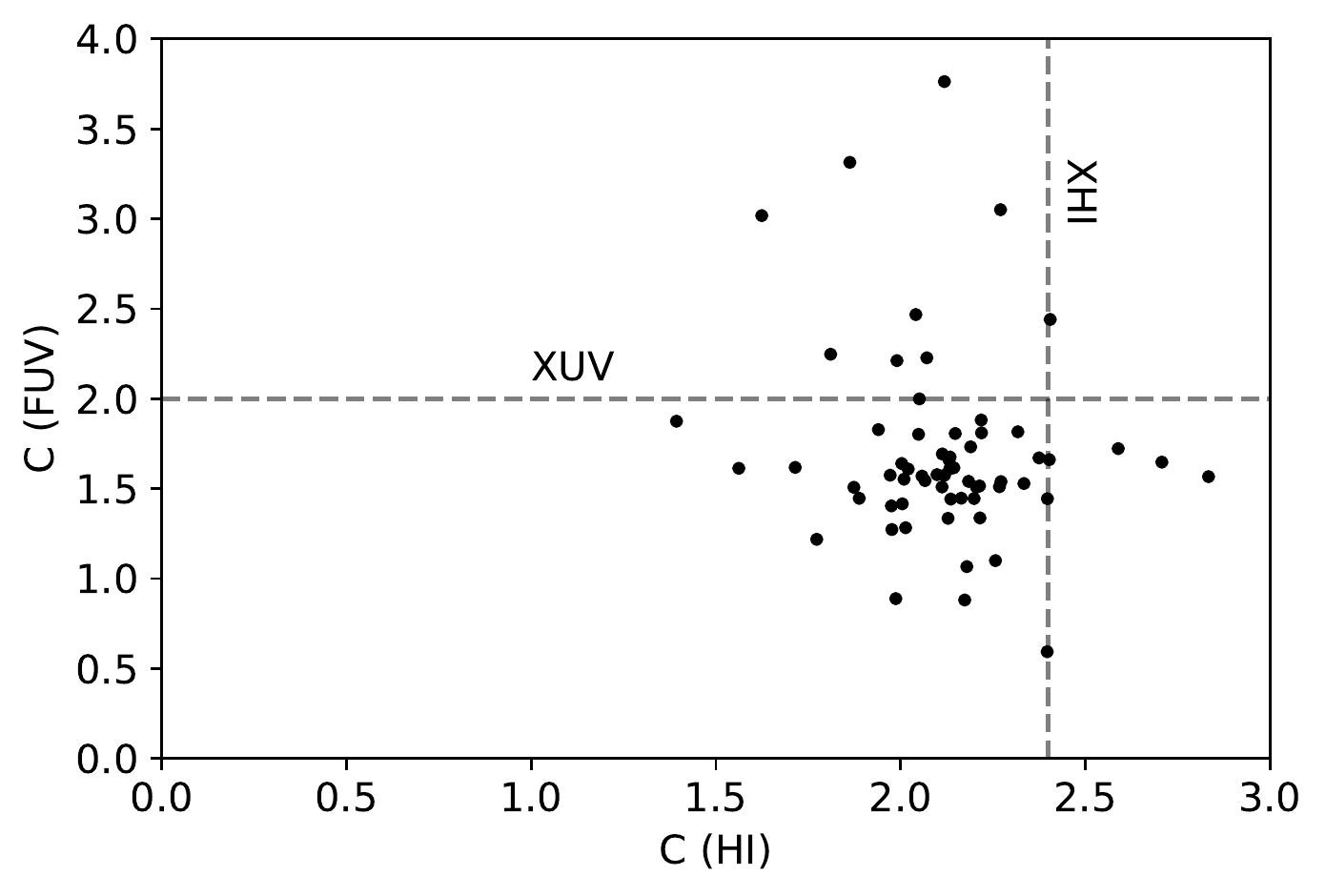}
    \caption{The concentration values for \HI{} and FUV for the overlap in samples. Extended \HI{} and UV disks potentially stand out by their high concentration value. The dashed lines are the criteria for extended sources from \protect\cite{Holwerda12c}. }
    \label{f:fuvhi:c}
\end{figure}{}

\subsection{Truncated and extended \HI{} disks}

\cite{Holwerda11e} noted that \HI{} computed concentration may be linked to \HI{} stripping and truncation. Figure \ref{f:fuvhi:c} shows that relatively few galaxies in the overlap sample exhibit the low values of \HI{} concentration ($C<1.5$) associated with truncated \HI{} disks. 
In the optical, the typical exponential disc value is $C=2-3$, a spheroidal galaxy much higher. 

We view these low concentration \HI{} disks as possibly truncated, similar to those observed by \cite{Chung09} for the Virgo cluster at slightly higher spatial resolution (15") with the VLA. \cite{Holwerda11e} found in the VIVA data that low-concentration \HI{} disks are more likely truncated. 
% Only one Hydra cluster galaxy observed has a low concentration (J104447-270553, $C_{HI} = 2.42$).

Conversely, high values of concentration could be an indication of an extended \HI{} or FUV disk.  Figure \ref{f:fuvhi:c} shows high values for mutually exclusive subsamples (listed in tables \ref{t:xhi} and \ref{t:xuv}). 
% \begin{figure}
%     \centering
%     \includegraphics[width=0.5\textwidth]{Figures/hi_R_C.pdf}
%     \caption{The distance from the center of the Hydra cluster and the \HI{} concentration. Three galaxies stand out with low \HI{} concentration which may point to pre-processing during a first cluster passage. }
%     \label{f:R:C}
% \end{figure}

% \begin{figure}
%     \centering
%     \includegraphics[width=0.5\textwidth]{Figures/hi_dproj_C.pdf}
%     \caption{The distance from the center of the Hydra cluster and the \HI{} concentration. Three galaxies stand out with low \HI{} concentration which may point to pre-processing during a first cluster passage. }
%     \label{f:R:C}
% \end{figure}{}

\begin{table}
    \centering
    \begin{tabular}{l l l l l l}
FUV & & & & \\
cat id      & $C_{FUV}$  & $A_{FUV}$ & $M_{20, FUV}$ & $C_{HI}$ & \\ 
\hline
\hline
J100720-262426 & 2.23 & -3.27 & -0.65 & 2.07 & *\\ 
J101035-254920 & 3.02 & -0.13 & -1.31 & 1.62 & \\ 
J101359-253824 & 2.44 & -0.15 & -0.87 & 2.40 & *\\ 
J101434-274133 & 3.31 &  0.11 & -1.40 & 1.86 & \\ 
J101945-272719 & 3.05 &  0.26 & -1.04 & 2.27 & *\\ 
J102207-282201 & 2.47 & -0.61 & -1.04 & 2.04 & *\\ 
J102413-284853 & 2.21 & -0.21 & -0.84 & 1.99 & \\ 
J102600-280334 & 3.76 &  0.18 & -0.73 & 2.12 & *\\ 
J102605-280710 & 2.25 & -1.92 & -0.79 & 1.81 & \\ 

% J104245-264738 & 2.23 & -3.27 & -0.65 & *\\
% J104059-270456 & 3.02 & -0.13 & -1.31 & *\\
% J103918-265030 & 2.44 & -0.15 & -0.87 & \\
% J103924-275442 & 3.31 & 0.11 & -1.40 & *\\
% J103749-270715 & 3.05 & 0.26 & -1.04 & *\\
% J103702-273359 & 2.47 & -0.61 & -1.04 & *\\
% J103645-281010 & 2.21 & -0.21 & -0.84 & \\
% J103521-274137 & 3.76 & 0.18 & -0.73 & \\
% J103507-275923 & 2.25 & -1.92 & -0.79 & *\\
\hline
NUV & & & & \\
cat id & $C_{NUV}$  & $A_{NUV}$ & $M_{20, NUV}$ & $C_{HI}$ & \\
\hline
J100720-262426 & 2.21 & -2.91 & -0.79 & 2.07 \\ 
J101035-254920 & 2.83 & 0.03 & -1.25 & 1.62 \\ 
J101359-253824 & 3.11 & -0.04 & -1.32 & 2.27 \\ 
J101434-274133 & 2.06 & -0.05 & -0.86 & 1.99 \\ 
J101945-272719 & 2.25 & -1.28 & -0.65 & 1.81 \\ 
% J104245-264738 & 2.21 & -2.91 & -0.79 & *\\
% J104059-270456 & 2.83 & 0.03 & -1.25 & *\\
% J103749-270715 & 3.11 & -0.04 & -1.32 & *\\
% J103645-281010 & 2.06 & -0.05 & -0.86 & *\\
% J103507-275923 & 2.25 & -1.28 & -0.65 & *\\
\hline
\hline
\end{tabular}
    \caption{The galaxies with high concentration values in FUV (Figure \ref{f:fuvhi:c} and NUV with their asymmetry and $M_{20}$ values. These are potential XUV disks in the Hydra cluster. Asterix indicates a candidate XUV disks based on the asymmetry and $M_{20}$ values.}
    \label{t:xuv}
\end{table}

\begin{table}
    \centering
    \begin{tabular}{c|c c c}
cat id & $C_{HI}$ & A & $M_{20}$\\ 
\hline
\hline
J100539-282633 & 2.47 & 0.40 & -1.29 \\ 
J100634-295615 & 2.52 & 0.36 & -1.52 \\ 
% J101208-252000 & 2.71 & 0.27 & -1.76 & 1.65 \ \ 
% J101247-275028 & 2.83 & 0.17 & -1.33 & 1.57 \ \ 
% J101359-253824 & 2.40 & 0.24 & -1.46 & 2.44 \ \ 
% J102059-265129 & 2.40 & 0.50 & -1.38 & 1.66 \ \ 
J102341-291347 & 2.45 & 0.39 & -1.42 \\ 
J102416-284343 & 2.41 & 0.46 & -1.56 \\ 
% J102447-264054 & 2.59 & 0.50 & -1.48 & 1.72 \ \ 
J103015-270743 & 2.55 & 0.39 & -1.63 \\ 
J103420-264728 & 3.02 & 0.41 & -1.82 \\ 
J103543-255954 & 2.49 & 0.63 & -1.28 \\ 
J103603-245430 & 2.43 & 0.32 & -1.61 \\ 
J103818-285023 & 2.71 & 0.35 & -1.62 \\ 
J103853-274100 & 2.80 & 0.63 & -1.48 \\ 
J103902-291255 & 2.43 & 0.30 & -1.50 \\ 
J103915-301757 & 2.53 & 0.21 & -1.09 \\ 
J104339-285157 & 2.76 & 0.24 & -1.59 \\ 
J104442-290119 & 2.89 & 0.39 & -1.61 \\ 
J104447-270553 & 2.42 & 0.43 & -1.48 \\ 
% J104442-290119 & 3.05 \\
% J104339-285157 & 2.81 \\
% J103722-273235 & 2.43 \\
% J103603-245430 & 2.51 \\
% J103539-284606 & 2.66 \\
% J103420-264728 & 3.06 \\
% J103322-271142 & 2.45 \\
% J103241-273137 & 2.41 \\
% J103244-283639 & 2.44 \\
\hline
\hline
\end{tabular}
    \caption{The galaxies with high concentration values in \HI{} (Figure \ref{f:fuvhi:c} but with no FUV or NUV counterpart. These are extended \HI{} disks that may hold XUV disks in the Hydra cluster.}
    \label{t:xhi}
\end{table}

\begin{figure*}
    \centering
    \includegraphics[width=\textwidth]{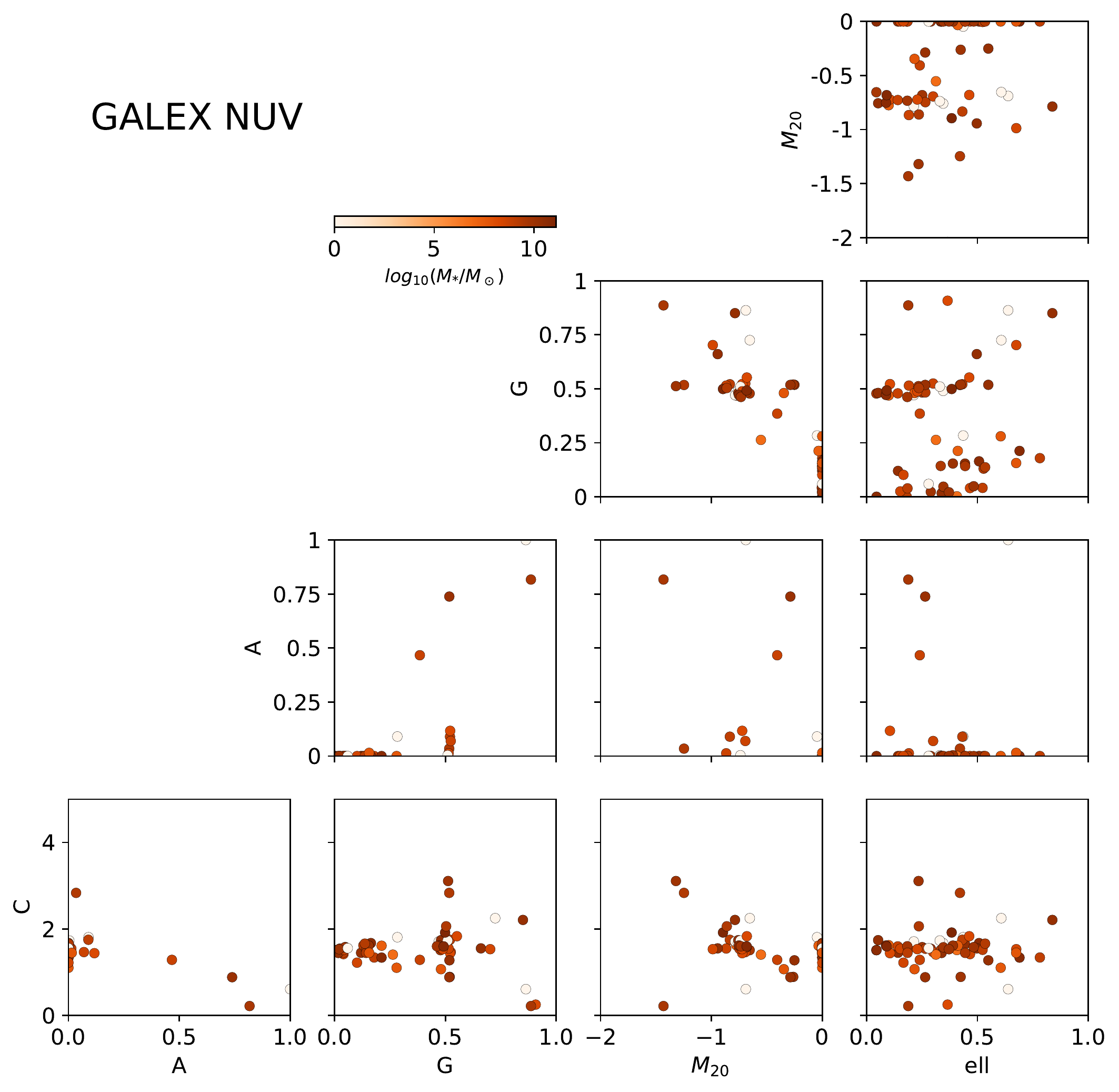}
    \caption{The distribution of NUV and FUV morphometric parameters for the objects with ultraviolet observations. There is NUV/FUV information on 30 objects (63 with catalog entries). Those with e.g. A=0 are the result of morphometrics on fixed pixel values of the area outside the FOV, they encode the \HI{} shape. The colour bar is the stellar mass of these galaxies.}
    \label{f:corner:uv}
\end{figure*}{}

\subsection{FUV Morphometrics}

Figure \ref{f:corner:uv} shows the corner plot for both FUV and NUV galaxies. Some of the objects included have unphysical values (e.g. A=0) and are likely included despite no flux in the \HI{} contour. In general, the NUV/FUV catalog is much smaller than the \HI{} catalog (30 objects in the GALEX fiels with 63 with nonzero GALEX morphometric values and 272 \HI{} morphometric values), calculated for all the objects in GALEX footprint.

\begin{figure*}
    \centering
    \includegraphics[width=0.49\textwidth]{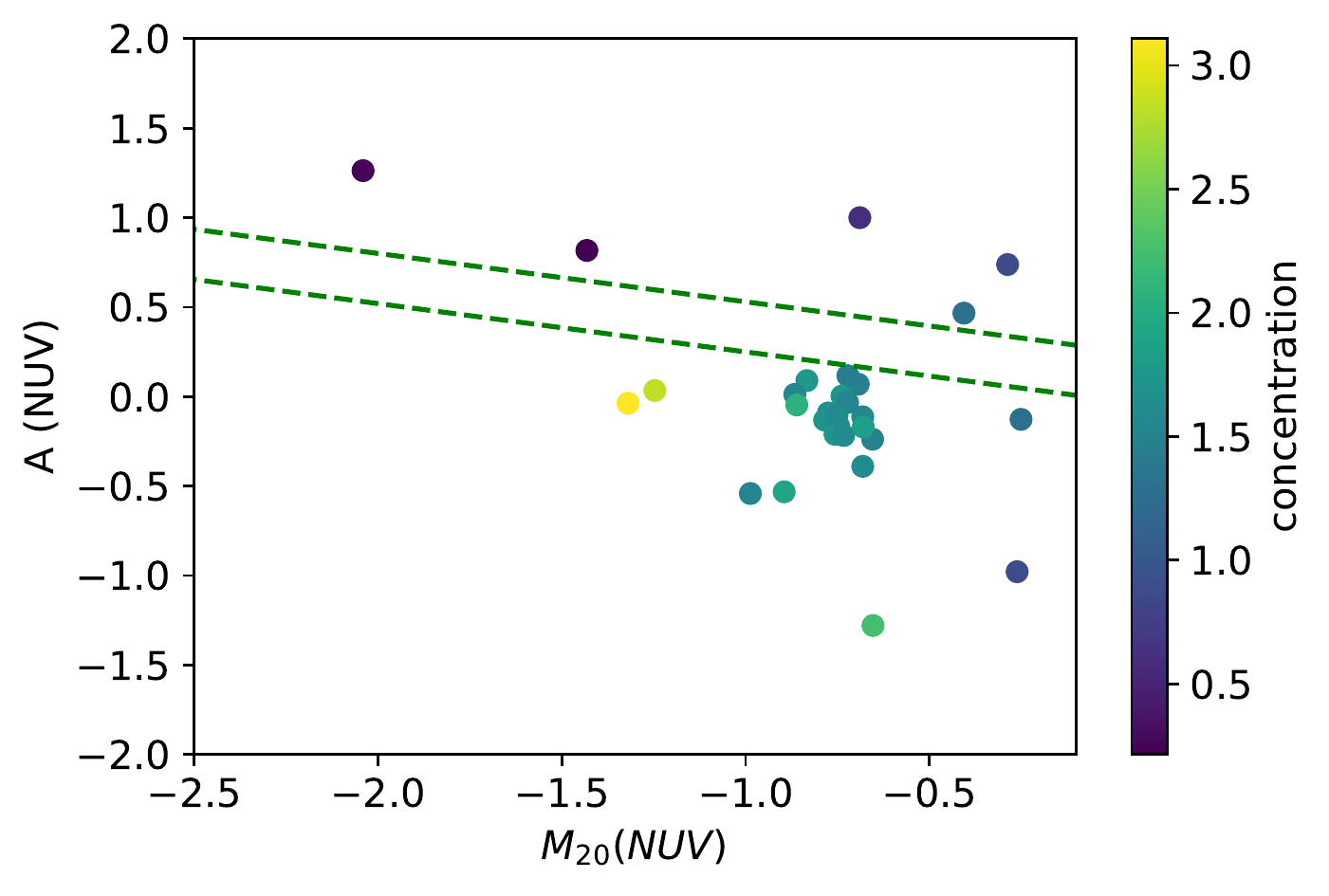}
    \includegraphics[width=0.49\textwidth]{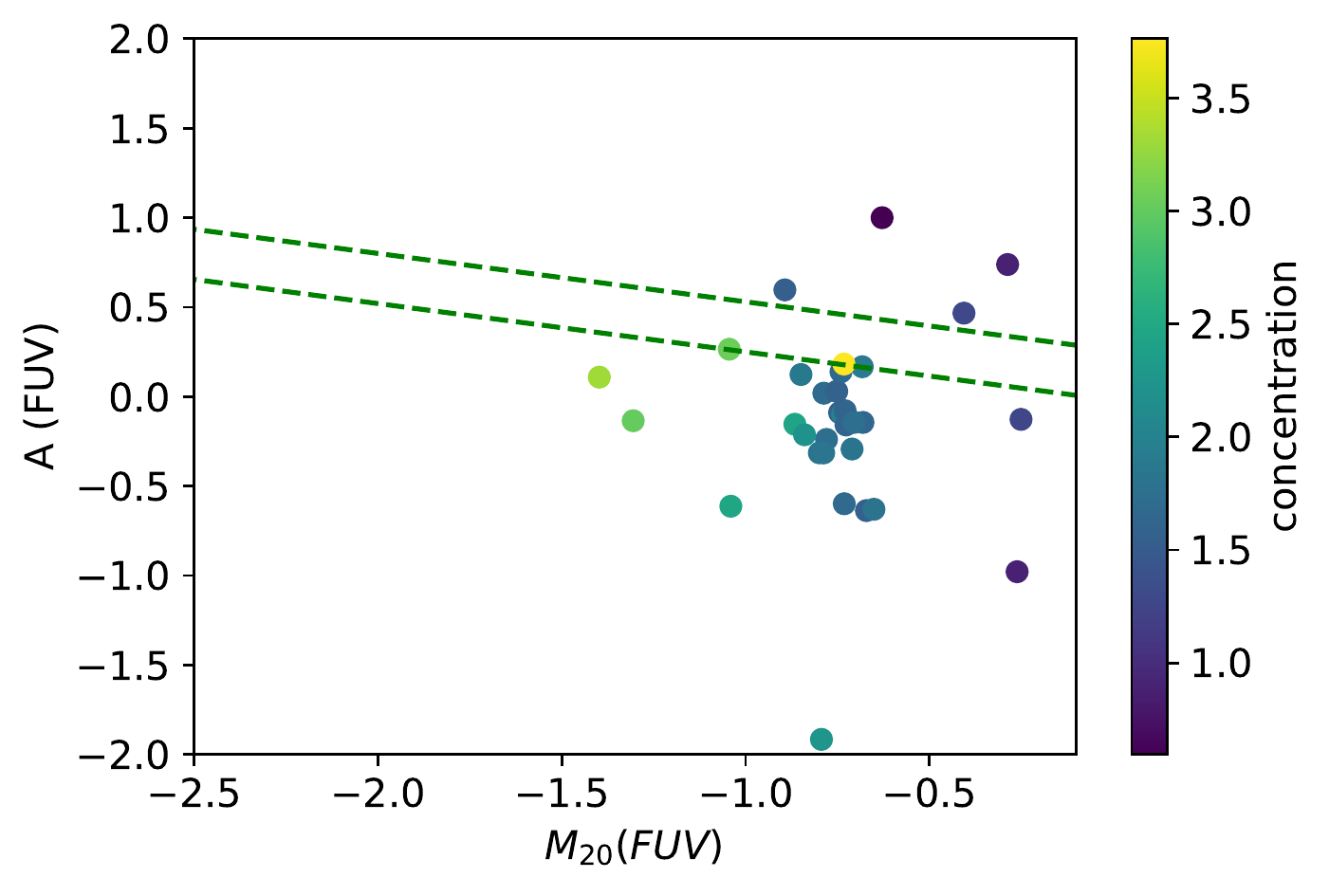}
    \caption{The asymmetry and $M_{20}$ parameters from {\sc statmorph} in the GALEX FUV and NUV channels with concentration colour-coded. The criterion identified by \protect\cite{Holwerda12c} as potentially useful to identify XUV disks is marked with green dashed lines. In between these lines were the known XUV disks identified in their WHISPS/GALEX catalog. 
    Galaxies with high values of concentration, low values of asymmetry and $M_{20}$ are good candidates for XUV disks. }
    \label{f:xuv}
\end{figure*}{}

\subsection{XUV disk candidates}

\cite{Holwerda12c} noted that the \HI{} outer contour is a good way to compute the UV morphology as it clearly delineates which UV clumps of emission belong to a disk galaxy and which do not. Figure \ref{f:xuv} shows the FUV $M_{20}$ and Asymmetry values with concentration colour-coded. If we use the $M_{20} - A$ criterion from \cite{Holwerda12c}, very few objects in the Hydra cluster fall under the XUV classification. However, the implementation of Asymmetry differs from \cite{Holwerda12c}. This emphasizes our point that for each survey, wavelength and implementation, new criteria will need to be calibrated to identify objects of interest. Low values of asymmetry and $M_{20}$ with high concentration values will be the candidates for XUV disks. \cite{Holwerda12c} noted that the fraction identified and specific criteria used to identify XUV disks depends on both \HI{} and UV depth and \HI{} resolution (e.g. WHISPS vs THINGS resolution)
Based on this initial identification, 5/32 galaxies ($\sim$15\%, see Table \ref{t:xuv}) in the Hydra cluster have an XUV disk, a substantial reduction (factor 5-6) compared to the field \citep{Lemonias11,Moffett12}.

\begin{figure*}
    \centering
    \includegraphics[width=0.49\textwidth]{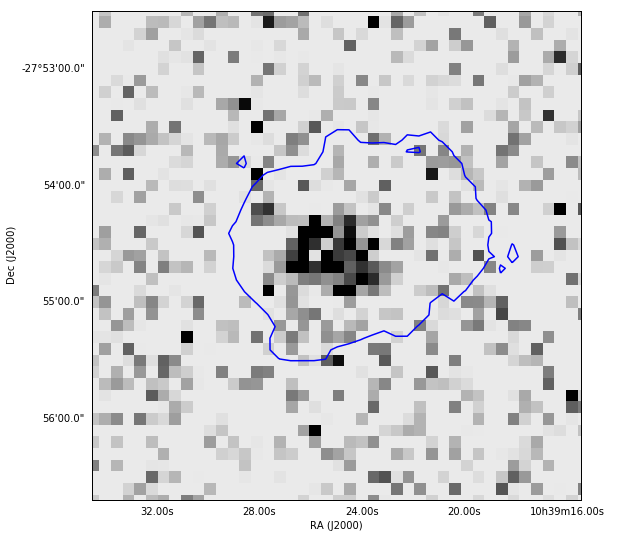}
    \includegraphics[width=0.49\textwidth]{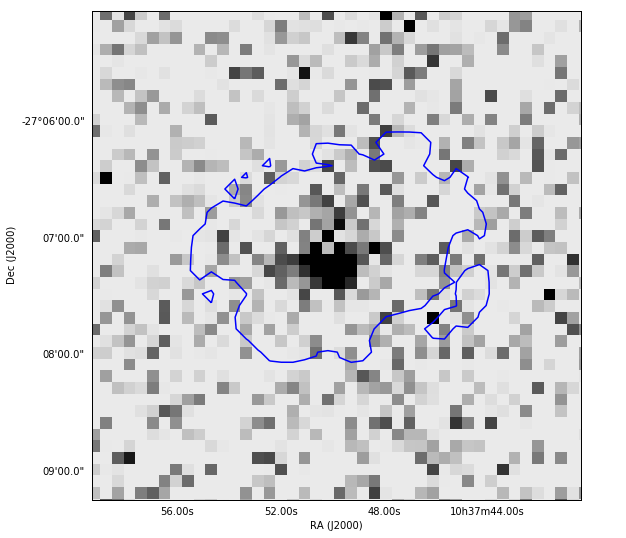}
    \includegraphics[width=0.49\textwidth]{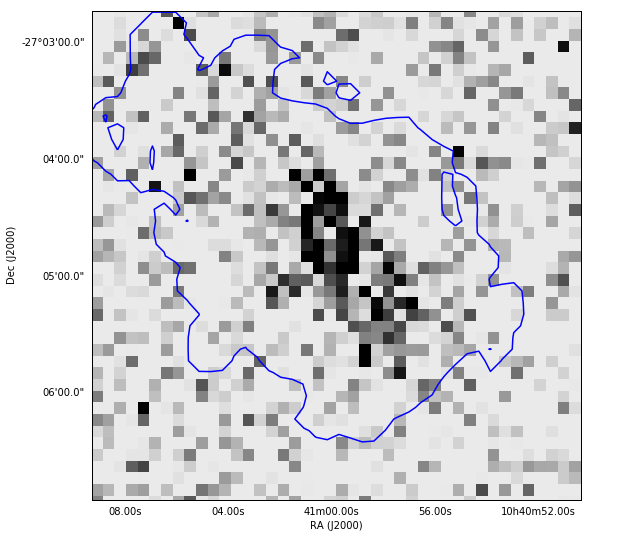}
    \includegraphics[width=0.49\textwidth]{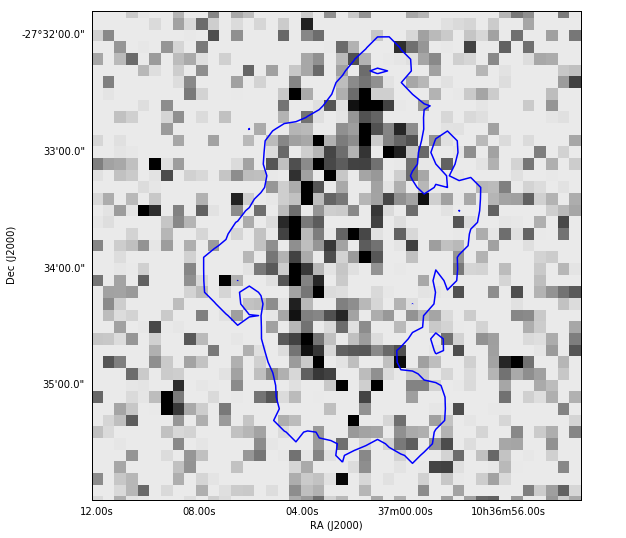}
    \caption{Examples of GALEX FUV cutouts with the \HI{} contour overlaid. These are reliable XUV detections with high UV values of concentration. The top examples could have been detected using an optical isophote but lower examples show the benefit of an \HI{} contour (10 Jy/beam$\times$Hz).           J103924-275442 (top-left), J103749-270715 (top-right), J104059-270456 (bottom-left), and J103702-273359 (bottom-right).}
    \label{f:fuv:examples}
\end{figure*}{}

Figure \ref{f:fuv:examples} shows a few examples of FUV images of high FUV concentration with the WALLABY \HI{} contour. It highlights how extended FUV sources can be found through computed apertures, but outer disk flux contributions will be missed without the \HI{} contour definition. Even with the most shallow GALEX observations typically available (AIP 200 s.) and greater distances than the resolved galaxies of THINGS and WHISP \citep{Holwerda12c}, one can identify extended UV disks through their morphometrics. 

\begin{figure*}
    \centering
    \includegraphics[width=\textwidth]{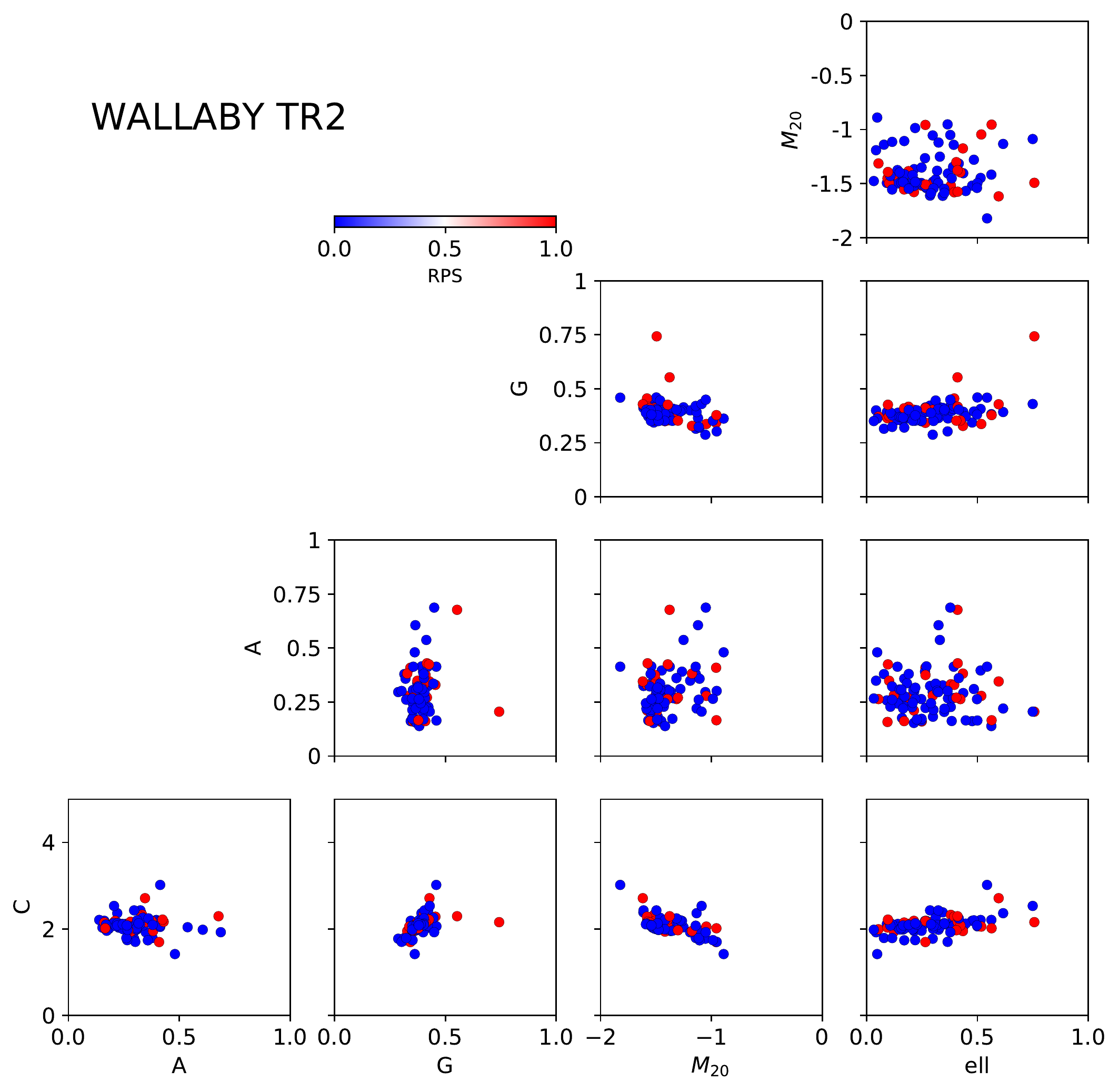}
    \caption{The C-A-G-M20-ellipticity  \HI{} morphometric feature space with the RPS flag from \protect\cite{Wang21} marked with the color: red galaxies are undergoing RPS.}
    \label{f:HImorph:flagRPS-1}
\end{figure*}

\begin{figure*}
    \centering    \includegraphics[width=\textwidth]{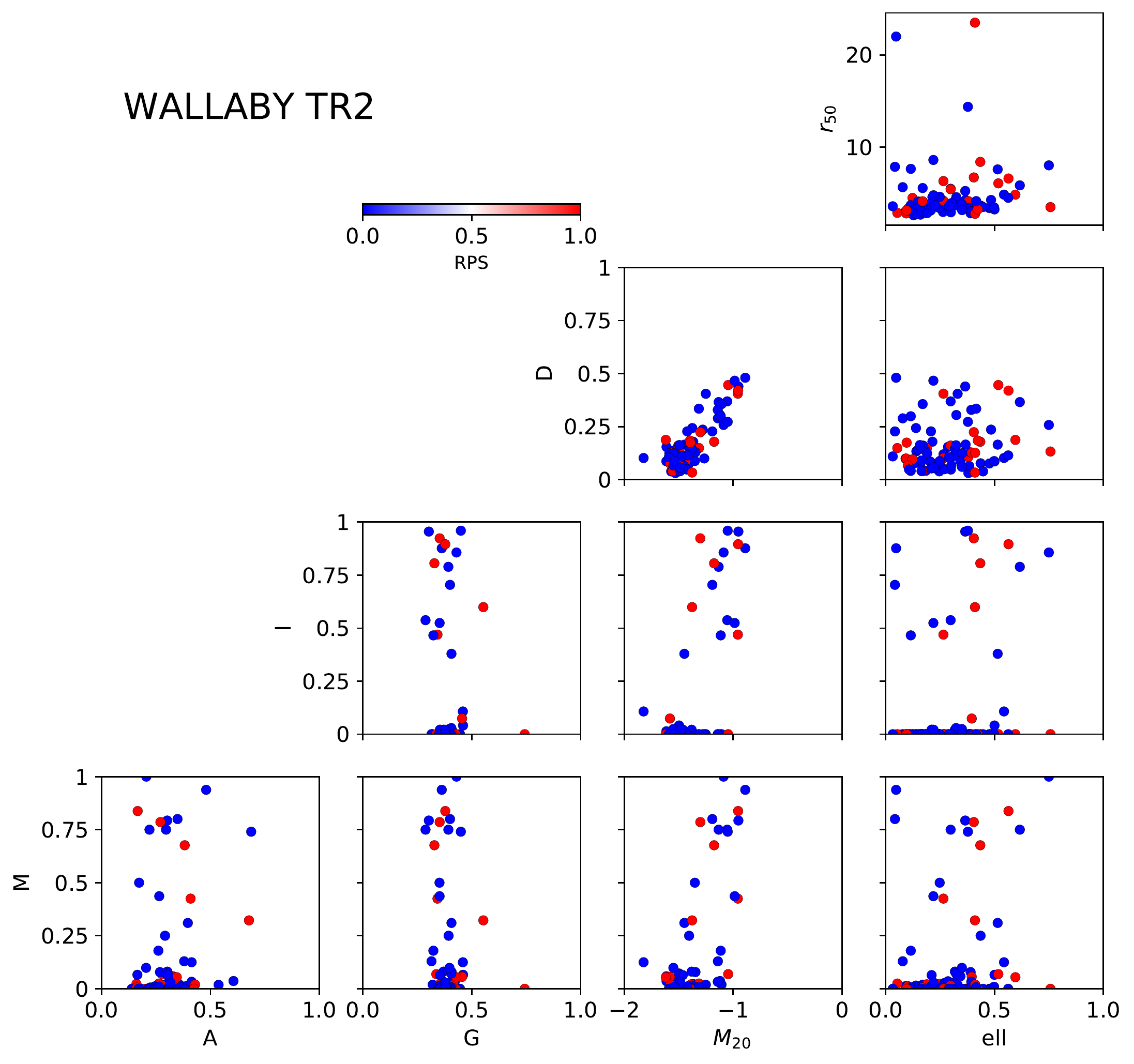}
    \caption{The  M-I-D-ell \HI{} morphometric feature space with the RPS flag from \protect\cite{Wang21} marked with the color: red galaxies are undergoing RPS.}
    \label{f:HImorph:flagRPS-2}
\end{figure*}

% \begin{figure*}
%     \centering
%     % \includegraphics[width=0.49\textwidth]{Figures/wally_hi_cagm20_corner_flagr1RPS.pdf}
%     % \includegraphics[width=0.49\textwidth]{Figures/wally_hi_MID_corner_flagr1RPS.pdf}
%     \includegraphics[width=0.49\textwidth]{PubFigures/holwerda22b_f10a.pdf}
%     \includegraphics[width=0.49\textwidth]{PubFigures/holwerda22b_f10b.pdf}
%     \caption{The C-A-G-M20-ellipticity (left) and the M-I-D-ell \HI{} morphometric feature space with the candidate resolved RPS flag (r1) from \protect\cite{Wang21} marked with the color: red galaxies are candidates for undergoing RPS.}
%     \label{f:HImorph:flagr1RPS}
% \end{figure*}

\section{Ram-pressure Stripping}
\label{s:RPS}

\cite{Wang21} present estimates of the ram-pressure stripping (RPS) that \HI{} galaxies in the Hydra cluster are undergoing based on the same data. They use a holistic approach to classify galaxies as undergoing RPS using \HI{} appearance, kinematics and stellar mass surface density. 
They flag galaxies undergoing RPS and candidate galaxies for undergoing RPS (RSP-r1). Figures \ref{f:HImorph:flagRPS-1} and \ref{f:HImorph:flagRPS-2} show the \HI{} morphometric feature space with the RPS flag and candidate RPS flag (RPS-r1) indicated. 

Both flags would be of interest for Machine Learning (hereafter ML) as one could use morphology to identify RPS without inspection or one could flag candidate RPS galaxies. In the \HI{} morphometrics (Figures \ref{f:HImorph:flagRPS-1} and \ref{f:HImorph:flagRPS-2}), the two populations are well-mixed, spanning the full ranges of values. It may well be that in the full feature space or a subset thereof, the two may well be separable. With no immediate clear separation in this parameter space, ML techniques developed to identify separation with a single hyperplane or a limited set of choices potentially could.

% the RPS appears separable, especially in the CAGM20 space 

% \begin{figure*}
%     \centering
%     \includegraphics[width=\textwidth]{PubFigures/holwerda22b_f11.png}
%     \caption{The decision tree to classify the $f_{RPS}$ from \protect\cite{Wang21} using the full \HI{} morphometric feature space. A depth of 13 is needed to split the sample, using all \HI{} morphometric features.}
%     \label{f:tree:fRPS:full}
% \end{figure*}

\section{Machine Learning Classification of RPS from \HI{} morphometrics}
\label{s:ML}

With a substantial feature space in the \HI{} morphology and no immediate intuition on where RPS galaxies should lie within this space, we attempted to use the \HI{} morphometric features to classify the RPS flags of \cite{Wang21} using different ML classifier from {\sc sklearn} \citep{sklearn}. The goal is to identify which \HI{} morphometric parameters are best in pre-selecting RPS,  what cuts in that space could lead to RPS candidates being identified in the remaining WALLABY survey, and how well ML can identify RPS in the \HI{} feature space.

Because the non-RPS galaxies outnumber the RPS galaxies in the Hydra catalogs, we balance the data using {\sc smote} in the {\sc imblearn} package which supplements the {\sc sklearn} one for machine learning. The sample is normalized (whitened) using {\sc sklearn}'s standard scaler. The sample is then split 80-20\% randomly for training and testing respectively (400/100 instances after {\sc smote}). This is a smaller sample than typical in machine learning applications. Our aim however is to evaluate the future applicability of the \HI{} morphometrics feature space to identify RPS.

We apply a decision tree, k-nearest neighbour, a support-vector machine, and a random forest classifier to estimate how well one can distinguish RPS galaxies in the \HI{} morphometric space. We calculate the accuracy (fraction of correct classifications), the precision (the fraction of true positives over true negatives), the recall (the fraction of true positives of total positives) and the F1 = (precision $\times$ recall) / (precision+recall). 
Before {\sc smote} to rebalance the data, the F1 scores were poor for all ML algorithms that follow.

\subsection{Decision Tree}

A decision tree classifier is a series of criteria in feature space that ultimately split the training sample according to the labels provided. We split our Hydra sample into 80\% training and 20\% test after {\sc smote} rebalancing. We trained on classifying either the RPS flag or the candidate RPS flag (RPS-r1) from \cite{Wang21}.

% \begin{figure*}
%     \centering
%     \includegraphics[width=\textwidth]{PubFigures/holwerda22b_f12.png}
%     \caption{The decision tree to classify the candidate RPS flag ($f_{r1,RPS}$) from \protect\cite{Wang21} using the full \HI{} morphometric feature space. A depth of 13 is needed to split the sample, using all \HI{} morphometric features.}
%     \label{f:tree:fr1RPS:cag20}
% \end{figure*}

\subsection{\HI{} Morphometrics Decision Tree}

% Figure \ref{f:tree:fRPS:full} and \ref{f:tree:fr1RPS:cag20} show two examples of the decision trees, one for the RPS flag and one for the candidate RPS flag (RPS-r1). 

We tried two decision trees, one for the RPS flag and one for the candidate RPS flag (RPS-r1). Both went to over a dozen forks in depth. 
Every permutation of the \HI{} morphometric feature space yielded a result (accuracy, precision, recall and F1) with the test sample. These are summarized in Table \ref{t:tree-performance}, showing the accuracy (combined precision for positive and negative) , precision (the fraction of selected positives correctly identified), recall (the fraction of positives correctly identified), and F1 ($\rm = precision \times recall / precision + recall$) metrics for each decision tree. Precision and recall are low for all trees, with Gini-R50 performing best for the RPS flag ($f_{RPS}$) and the C-A-G-M20 space for the candidate RPS flag ($f_{r1,RPS}$).
 
At present best, the \HI{} morphometric feature space can identify about 80\% of the galaxies undergoing RPS \citep[as identified by][]{Wang21}. Similar to what we found in \cite{Holwerda11e}, there are multiple processes influencing the \HI{} appearance and perhaps this feature space cannot do much better. However, the sample is still small compared to typical ML applications. 

\begin{table}
    \centering
    \begin{tabular}{l|l l l l}
            & Accuracy & Precision & Recall & F1 \\
\hline
\hline
$f_{RPS}$   &           &           &       &       \\

% G-R50       & 0.75 & 0.13 & 0.25 & 0.08 \\
HI C-A-G-M20-M-I-D-R50 & 0.80 & 0.71 & 0.89 & 0.39 \\
HI C-A-G-M20 & 0.80 & 0.84 & 0.86 & 0.43 \\
HI MID-R50 & 0.80 & 0.73 & 0.82 & 0.39 \\

\hline
$f_{r1,RPS}$   &           &           &       &       \\
HI C-A-G-M20-M-I-D-R50 & 0.83 & 0.85 & 0.82 & 0.42 \\
HI C-A-G-M20 & 0.78 & 0.80 & 0.80 & 0.40 \\
HI MID-R50 & 0.80 & 0.80 & 0.82 & 0.41 \\

\hline
\hline
    \end{tabular}
    \caption{The Accuracy, Precision, Recall and F1 metrics of the decision tree based on different feature choices to identify the unresolved and resolved ram-pressure stripping flags from \protect\cite{Wang21}.}
    \label{t:tree-performance}
\end{table}

% \subsection{FUV Morphometric Decision Tree}

% Lastly, we attempt to classify the RPS using the FUV morphometric feature space. This worked slightly better, but we caution that this is a smaller sample. 
% This feature space identifies candidate RPS galaxies with high precision (75\%) but still low recall (33\%) (Table \ref{t:tree-performance}). Figure \ref{f:tree:fuv} shows the best performing decision tree which is still seven branches deep.
% \begin{figure*}
%     \centering
%     \includegraphics[width=\textwidth]{Figures/tree_fuv_cagr50.pdf}
%     \caption{The decision tree based on FUV morphometrics identifying candidate RPS galaxies. The tree has high precision (75\%) but still low recall (33\%).}
%     \label{f:tree:fuv}
% \end{figure*}

\begin{table}
    \centering
    \begin{tabular}{l|l l l l}
            & Accuracy & Precision & Recall & F1 \\
\hline
\hline
$f_{RPS}$   &           &           &       &       \\
HI C-A-G-M20-M-I-D-R50 & 0.76 & 0.65 & 1.00 & 0.39 \\
HI C-A-G-M20 & 0.81 & 0.72 & 0.93 & 0.41 \\
HI M-I-D-R50 & 0.74 & 0.64 & 0.93 & 0.38 \\
\hline
$f_{r1,RPS}$   &           &           &       &       \\
HI C-A-G-M20-M-I-D-R50 & 0.80 & 0.75 & 0.92 & 0.41 \\
HI C-A-G-M20 & 0.69 & 0.67 & 0.84 & 0.37 \\
HI M-I-D-R50 & 0.74 & 0.74 & 0.80 & 0.38 \\
\hline
\hline
    \end{tabular}
    \caption{The Accuracy, Precision, Recall and F1 metrics of the kNN based on different feature choices to identify the unresolved and resolved ram-pressure stripping flags from \protect\cite{Wang21}.}
    \label{t:kNN-performance}
\end{table}

\subsection{K-Nearest Neighbour on \HI{} Morphometrics}

K-Nearest Neighbour classifies using the N nearest neighbours in the feature space. Table \ref{t:kNN-performance} shows the metrics for the RPS flag in different \HI{} morphometrics with N=3. Performance is similar or poorer than the decision trees; the precision is typically worse, the recall better, resulting in comparable F1 metrics. KNN classifiers works best in coherent, homogeneous and isometric areas within the feature space. If the labels are sharply split in the feature space, this approach has less applicability.

\subsection{Support Vector Machine on \HI{} Morphometrics}

Support Vector Machine (SVM) is a machine learning algorithm optimized to disentangle labeled populations within a feature space. The SVM algorithm is to find a hyperplane in an N-dimensional space, where N is the number of features, that distinctly classifies the data points according to their labels (in this case RPS or not). We use the {\sc sklearn} implementation of SVM to try to classify using the \HI{} morphometrics.

\begin{figure*}
    \centering
    \includegraphics[width=0.32\textwidth]{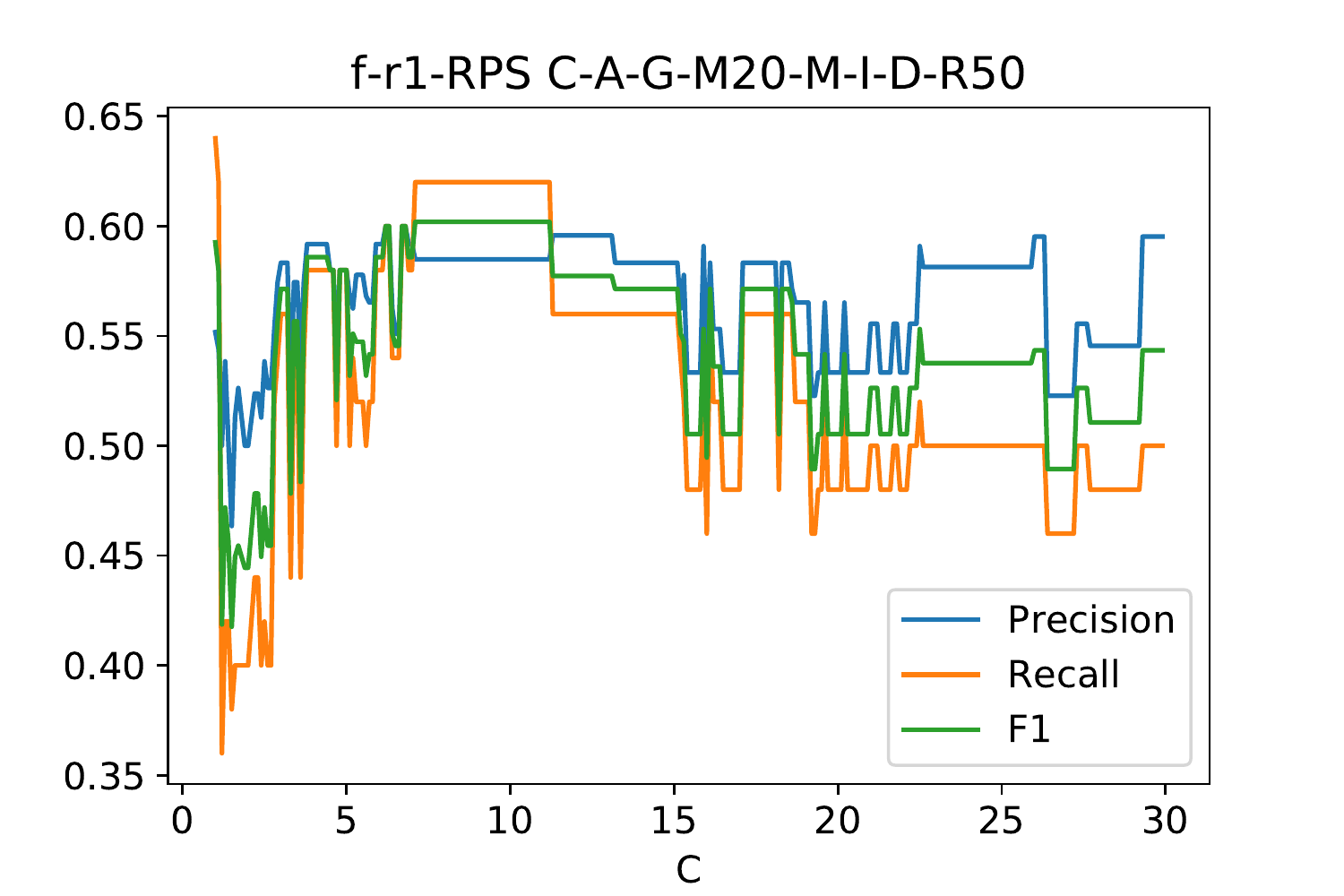}
    \includegraphics[width=0.32\textwidth]{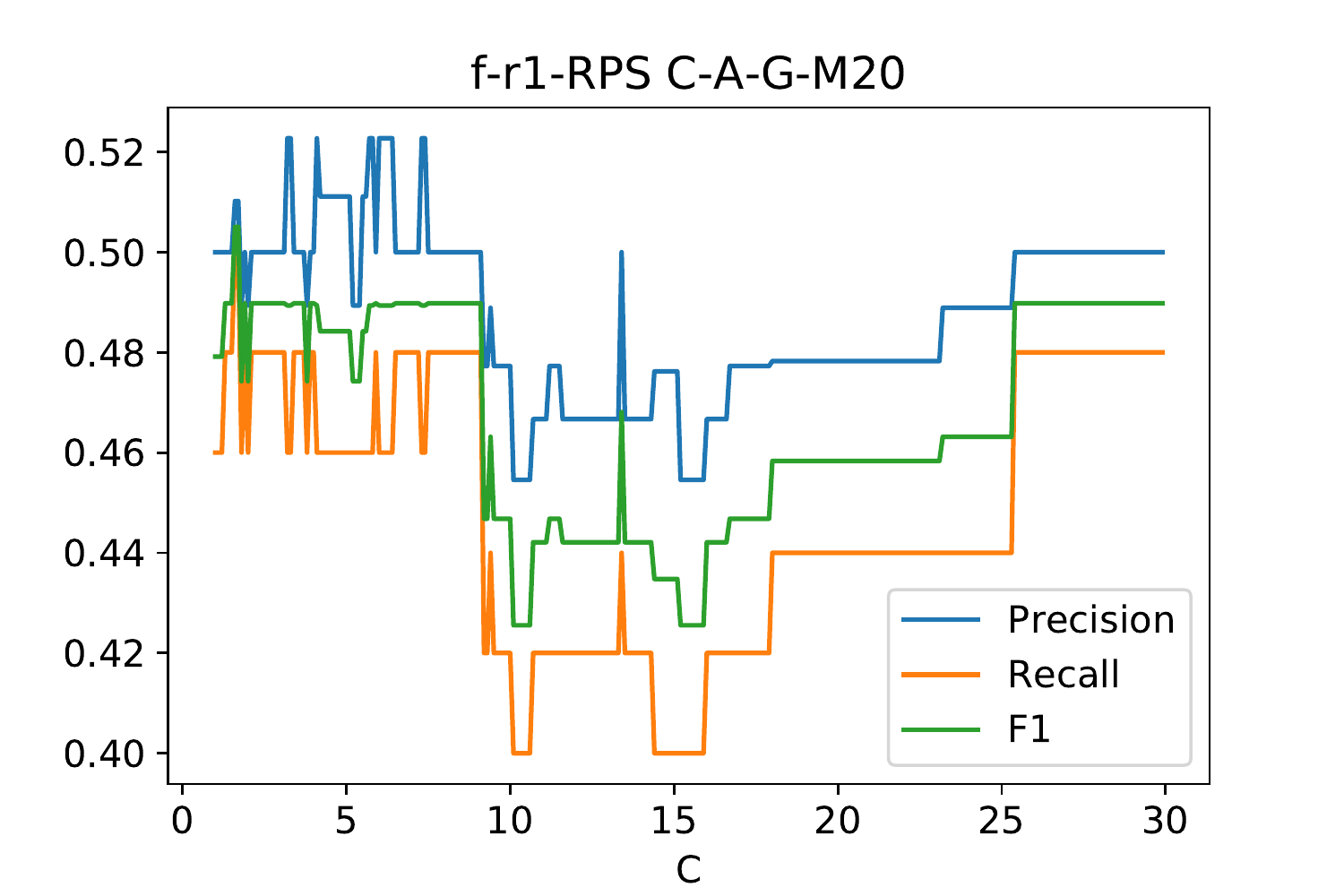}
    \includegraphics[width=0.32\textwidth]{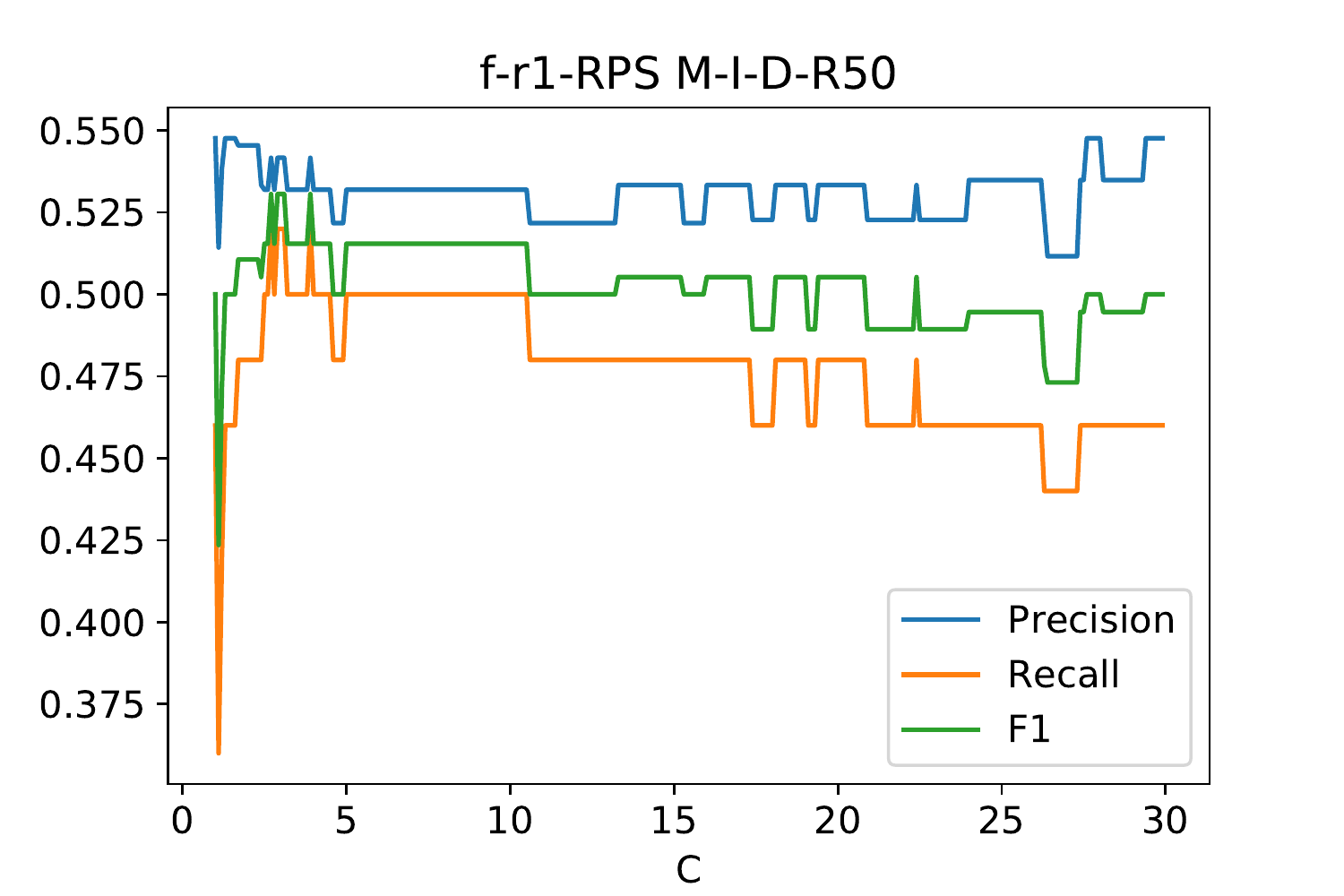}
    \caption{The metrics (Precision, recall and F1) as a function of the regularization parameter $C_{reg}$ of the SVM for three instances of the feature space: CAGM20MIDR50, CAGM20, and MIDR50 for the r1-RPS flag from \protect\cite{Wang21}.SVM performs poorly in separating RPS from non-RPS galaxies with every configuration of the feature space and regularization parameter $C_{reg}$.}
    \label{f:svm:C}
\end{figure*}

The main hyper-parameters regulating SVMs are the regularization parameter $C_{reg}$ and the choice and tuning of a convolution kernel. We opt for ``sigmoid" kernel, the default, and a ``polynomial" one, and plot the metrics as a function of the regularization parameter in figure \ref{f:svm:C}. Most combinations of feature space perform optimally around $C\sim5$. 
%For the full feature there is a high-precision, low-recall combination at C=2, which is a ``reject almost everything" scenario.
The metrics for $C_{reg}=5$ and the ``sigmoid'' kernel for each different feature set are listed in Table \ref{t:svm}. While the accuracy initially looks reasonably promising, both precision and recall leave much to be desired (neither realistically top 75\%). We note that the \HI{} morphometrics catalog is still small for machine learning applications with 272 total entries and 148 with RPS classifications and there is room for future improvements with larger samples.

Changing to a different kernel (from sigmoid to polynomial) does not improve performance. 
% Table \ref{s:svm2}

\begin{table}
    \centering
    \begin{tabular}{l l l l l}
features & Accuracy & Precision & Recall & F1\\
\hline
\hline
$f_{RPS}$ & & & & \\
\hline
% C-A-G-M20-M-I-D-R50 & 0.88 & 0.00 & 0.00 &  \dots  \\
% C-A-G-M20 & 0.85 & 0.33 & 0.50 & 0.20 \\
% M-I-D-R50 & 0.88 & 0.00 & 0.00 &  \dots  \\
% G-R50 & 0.88 & 0.00 & 0.00 &  \dots  \\
C-A-G-M20-M-I-D-R50 & 0.74 & 0.66 & 0.86 & 0.37 \\
C-A-G-M20 & 0.74 & 0.67 & 0.80 & 0.36 \\
M-I-D-R50 & 0.64 & 0.56 & 0.89 & 0.34 \\
\hline
\hline
$f_{r1,RPS}$ & & & & \\
\hline
% C-A-G-M20-M-I-D-R50 & 0.75 & 0.33 & 0.25 & 0.14 \\
% C-A-G-M20 & 0.80 & 0.00 & 0.00 &  \dots \\
% M-I-D-R50 & 0.78 & 0.43 & 0.38 & 0.20 \\
% G-R50 & 0.75 & 0.38 & 0.38 & 0.19 \\
C-A-G-M20-M-I-D-R50 & 0.61 & 0.63 & 0.66 & 0.32 \\
C-A-G-M20 & 0.60 & 0.63 & 0.64 & 0.32 \\
M-I-D-R50 & 0.49 & 0.62 & 0.16 & 0.13 \\
\hline
\hline
    \end{tabular}
    \caption{The SVM performance for different sub-sets of the feature space with fixed $C_{reg}=5$. $F1 = 2\times(\rm Recall \times Precision) / (Recall + Precision)$ is the weighted average of precision and recall. }
    \label{t:svm}
\end{table}

\begin{table}
    \centering
    \begin{tabular}{l l l l l}
features & Accuracy & Precision & Recall & F1\\
\hline
\hline
$f_{RPS}$ & & & & \\
\hline
C-A-G-M20-M-I-D-R50 & 0.52 & 0.48 & 1.00 & 0.32 \\
C-A-G-M20           & 0.51 & 0.47 & 1.00 & 0.32 \\
M-I-D-R50           & 0.51 & 0.47 & 1.00 & 0.32 \\
\hline
\hline
$f_{r1,RPS}$ & & & & \\
\hline
C-A-G-M20-M-I-D-R50 & 0.52 & 0.73 & 0.16 & 0.13 \\
C-A-G-M20           & 0.56 & 0.55 & 0.94 & 0.35 \\
M-I-D-R50           & 0.52 & 0.73 & 0.16 & 0.13 \\
\hline
\hline
    \end{tabular}
    \caption{The SVM performance for different sub-sets of the feature space with fixed $C_{reg}=5$ and a polynomial kernel convolution.} \label{t:svm2}
\end{table}

% \begin{figure*}
%     \centering
%     \includegraphics[width=0.49\textwidth]{Figures/fRPS_cagm20midr50_Cplot.pdf}
%     \includegraphics[width=0.49\textwidth]{Figures/fr1RPS_cagm20midr50_Cplot.pdf}
%     \includegraphics[width=0.49\textwidth]{Figures/fRPS_cagm20_Cplot.pdf}
%     \includegraphics[width=0.49\textwidth]{Figures/fr1RPS_cagm20_Cplot.pdf}
%     % \includegraphics[width=0.49\textwidth]{Figures/fRPS_midr50_Cplot.pdf}
%     \includegraphics[width=0.49\textwidth]{Figures/fRPS_gr50_Cplot.pdf}
%     % \includegraphics[width=0.49\textwidth]{Figures/fr1RPS_midr50_Cplot.pdf}
%     \includegraphics[width=0.49\textwidth]{Figures/fr1RPS_gr50_Cplot.pdf}
%     \caption{The metrics dependence on the regularization parameter C.}
%     \label{f:svm:C}
% \end{figure*}

% Figures_v3/RFC_cagm20midr50_matrix & 0.84 & 0.74 & 0.98 & 0.42 \\
% Figures_v3/RFC_cagm20_matrix & 0.86 & 0.79 & 0.93 & 0.43 \\
% Figures_v3/RFC_midr50_matrix & 0.71 & 0.62 & 0.91 & 0.37 \\
% Figures_v3/RFC_cagm20midr50_r1_matrix & 0.78 & 0.78 & 0.84 & 0.40 \\
% Figures_v3/RFC_cagm20_r1_matrix & 0.74 & 0.72 & 0.84 & 0.39 \\
% Figures_v3/RFC_midr50_r1_matrix & 0.77 & 0.78 & 0.80 & 0.40 \\
\subsection{Random Forest}

A random forests is an ensemble machine learning method for classification that works by constructing a multiple decision trees at training time into an ensemble (trees combined into a forest). For classification, the output of the random forest is the class selected by most trees. It has the benefit of more robustness and less sensitivity to outliers in the training set. 
For our catalog classification, this is the final option before one resorts to classifications with the images as inputs with e.g. convolutional neural networks etc.

Table \ref{t:rf} shows the performance of the random forest for different choices of the parameter space. It performs quite well with rebalancing the training data. Similar results on classification based on these morphometrics for simulated JWST images came to a similar conclusion that morphometrics and a random forest are well paired (Rose et. al. \textit{in preparation}). 

\begin{table}
    \centering
    \begin{tabular}{l l l l l}
features & Accuracy & Precision & Recall & F1\\
\hline
\hline
$f_{RPS}$ & & & & \\
\hline
C-A-G-M20-M-I-D-R50 & 0.84 & 0.74 & 0.98 & 0.42 \\
C-A-G-M20           & 0.86 & 0.79 & 0.93 & 0.43 \\
M-I-D-R50           & 0.71 & 0.62 & 0.91 & 0.37 \\
\hline
\hline
$f_{r1,RPS}$ & & & & \\
\hline
C-A-G-M20-M-I-D-R50 & 0.78 & 0.78 & 0.84 & 0.40 \\
C-A-G-M20           & 0.74 & 0.72 & 0.84 & 0.39 \\
M-I-D-R50           & 0.77 & 0.78 & 0.80 & 0.40 \\
\hline
\hline
    \end{tabular}
    \caption{The random forest performance for different sub-sets of the feature space.} \label{t:rf}
\end{table}

\begin{figure*}
    \centering
    \includegraphics[width=0.49\textwidth]{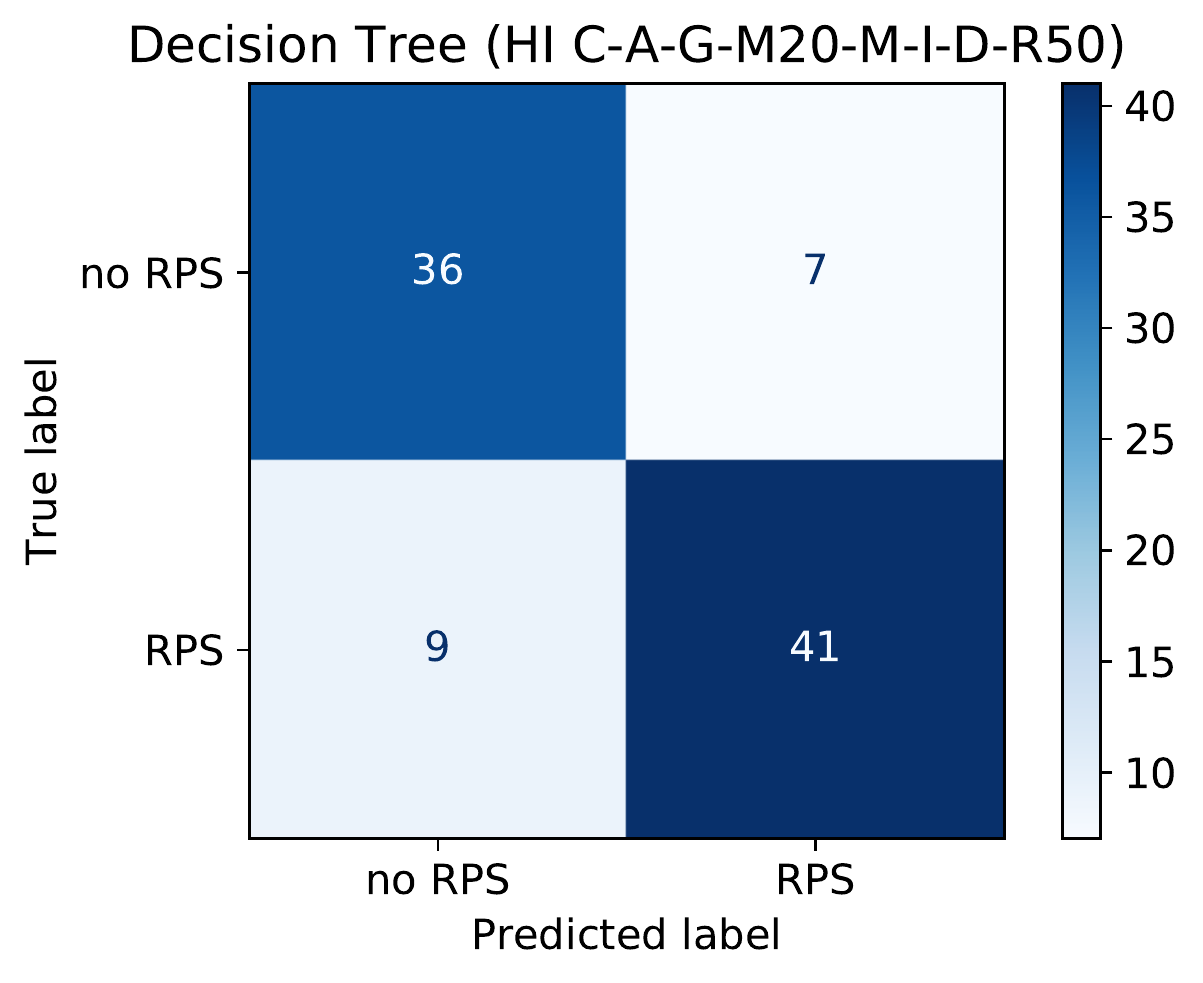}
    \includegraphics[width=0.49\textwidth]{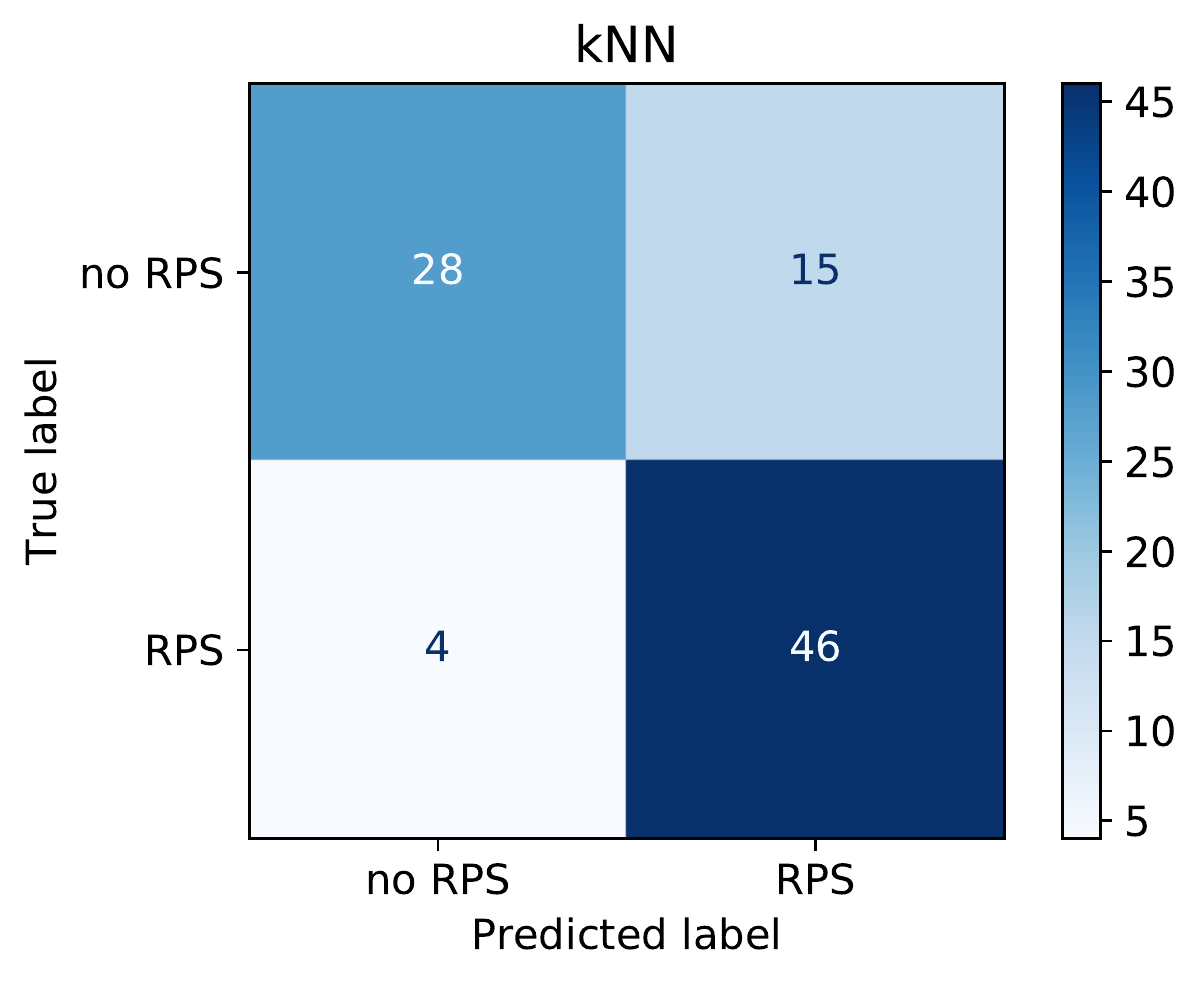}
    \includegraphics[width=0.49\textwidth]{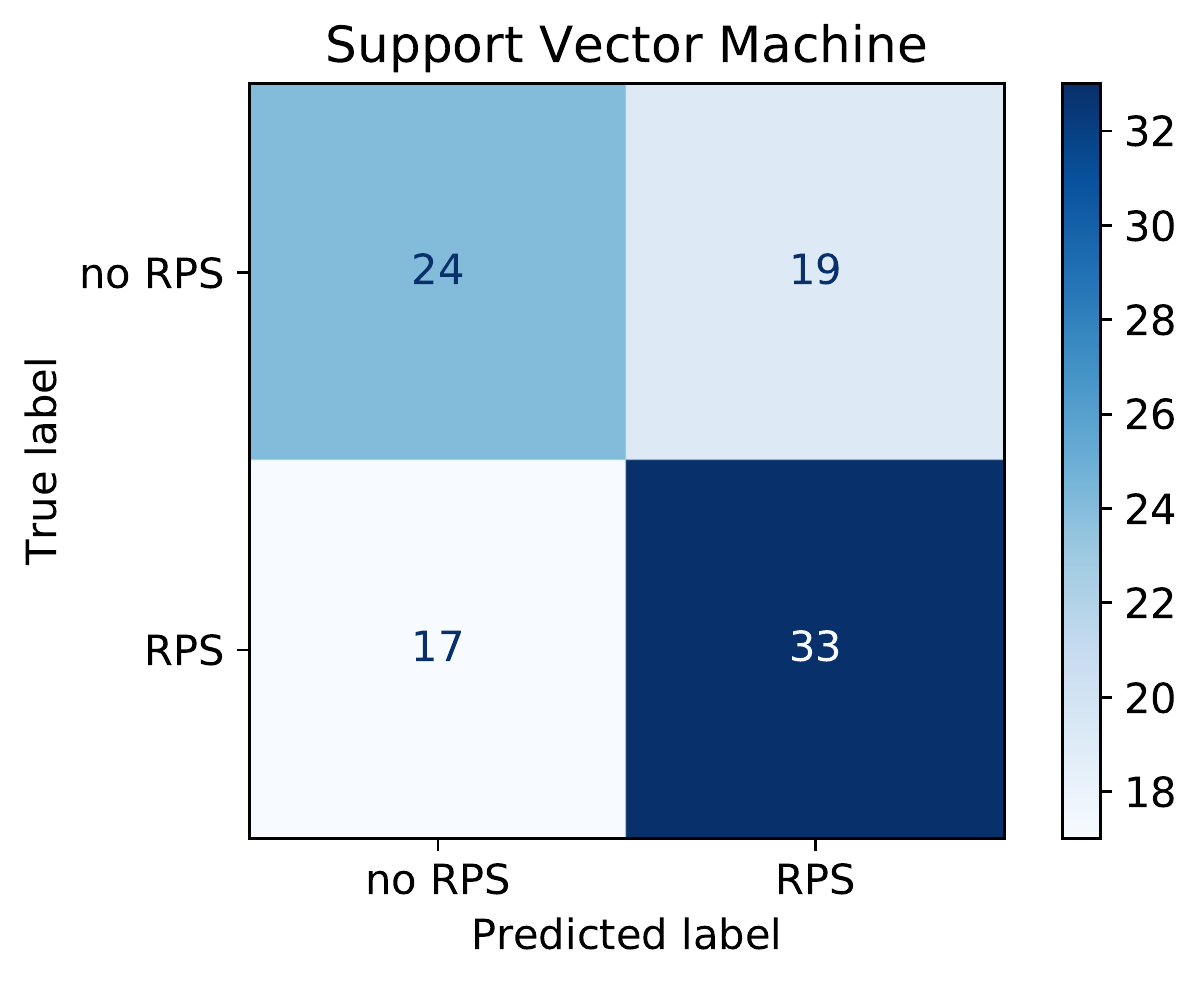}
    \includegraphics[width=0.49\textwidth]{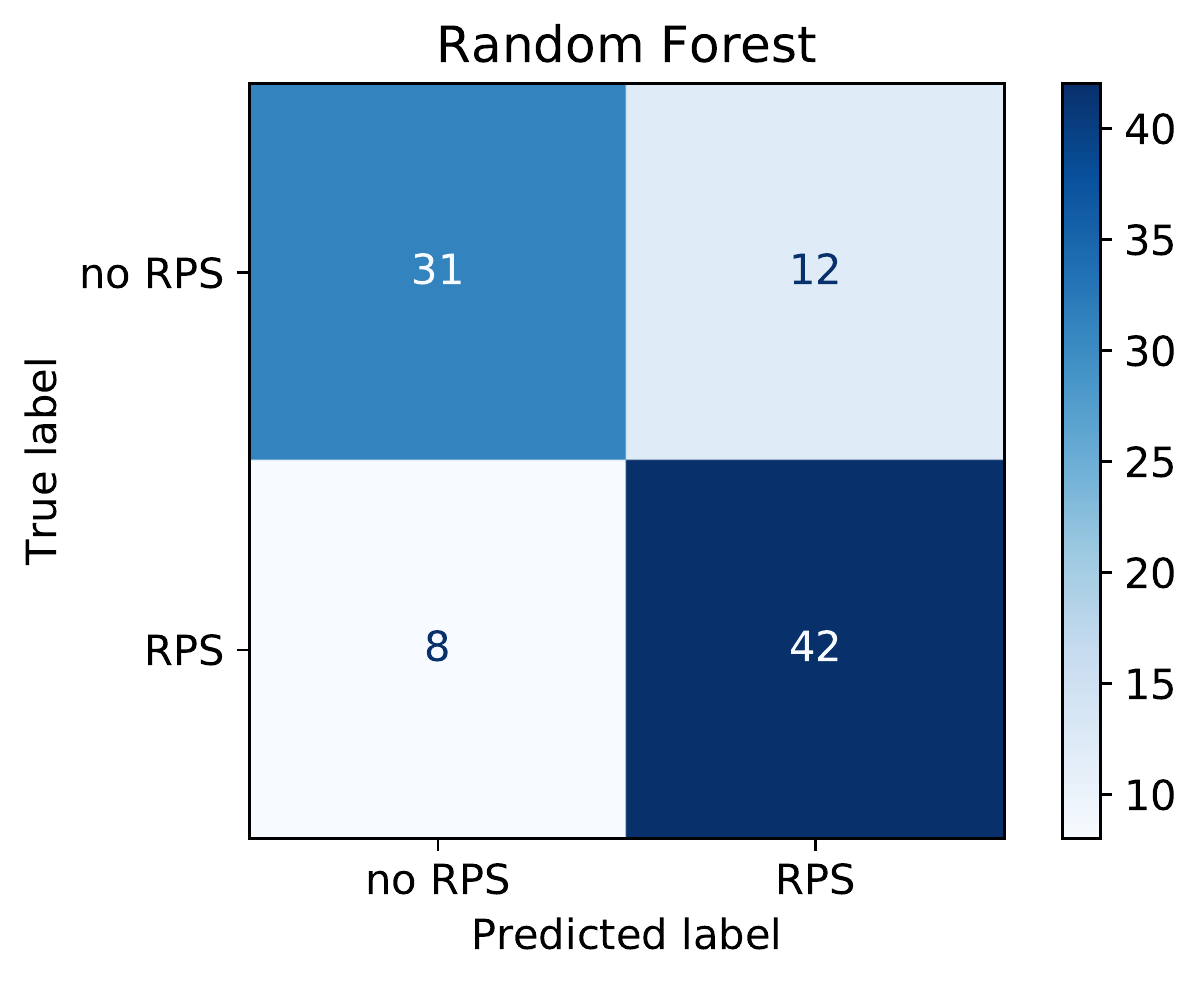}
    \caption{The classification matrices for the full \HI{} feature space (C-A-G-M20-M-I-D-R50) and the r1-RPS flag from \protect\cite{Wang21} for the decistion tree, kNN, SVM and RF.}
    \label{f:matrices}
\end{figure*}

\section{Concluding Remarks}
\label{s:conclusions}

The initial quality of the Hydra cluster WALLABY field is already enough to produce reliable morphology maps from the moment-0 (intensity) maps. We combined the \HI{} information with GALEX imaging for 30 out of the 272 galaxies in the field. In this way, the \HI{} information allows one to define the complete disk of galaxies, useful to identify which UV sources should be considered part of the galaxy and which are background sources \citep[cf the discussion of outermost \ion{H}{ii} regions in][]{Hunter18}.

%XUV
The fraction of XUV disks in the Universe is a potential constraint on how many disk galaxies are still accreting gas to their outskirts and how many are not. The outermost disk remains an excellent edge-case for star-formation in low-density ISM \citep{Watson16}, a potential way to rejuvenate spheroidal galaxies \citep{Moffett12}, and for tracing the chemical enrichment of disk galaxies \citep{Lopez-Sanchez15}.

Structural searches for XUV galaxy disks by \cite{Thilker07b} and \cite{Lemonias11} find that some 20--30\% of spirals have an XUV disk and 40\% of S0s \citep{Moffett12}, making this type of disk common but not typical for spiral and S0 galaxies. This is based mostly on local Universe data and on visual inspection \citep[see also][]{Thilker16} or a comparison with Spitzer radial profiles in deep imaging \citep{Bouquin15,Bouquin16,Bouquin18}.
However, with WALLABY and GALEX information combined, FUV morphometrics within the \HI{} contour poses a viable way to identify these XUV disks (5/32, 15\%). 
The fraction of XUV disks is lower than in the field, typically $\sim$40\%. 
The prerequisites are ASKAP \HI{} and GALEX FUV imaging, where even the shallowest GALEX exposures are sufficient thanks to the accurate delineation using the outer \HI{} contour using ASKAP. 

% HI Morphometrics
We computed the morphometrics over the moment-0 maps for all 272 galaxies in the Hydra TR2 catalog to act as the feature space to explore. A cluster population of galaxies at comparable distances with similar signal-to-noise limits on the observations is an ideal sample 

%RPS
Ram-pressure stripping is considered a principal avenue for galaxy transformation, in addition to tidal interactions, especially in dense environments. Identifying galaxies that are potentially undergoing ram-pressure stripping is therefore a motivator for \HI{} (morphological) studies. We compare our morphometric feature space to the RPS labels from \cite{Wang21} and find that Machine Learning classifiers --decision trees, Support Vector Machines, K-Nearest Neighbour, Random Forest-- do not fully separate RPS galaxies in \HI{} morphometric space yet.
Decision trees are impractically deep, the SVM struggle with precision but this problem is slightly better with kNN and RF. On the whole performance is middling (e.g. $\sim$0.6 accuracy, precision and recall for SVM) to fairly decent ($\sim$0.8 for the RF and kNN). On the whole, it points to reasonably good performance (80\% precision and recall) is possible for populations and acceptable performance for individual galaxies i.e. a ML algorithm can give a probability it is undergoing ram pressure stripping) with a larger training set. 

%The resulting metrics indicate that RPS and non-RPS galaxies are well mixed in morphometrics space. 
% The issue remains that a fraction of the RPS galaxies (20\%) are missed by any machine learning classifier using \HI{} morphometrics as the input. 
% HI
The feature space of \HI{} morphometrics used is either the full, (C-A-G-M20-M-I-D-R50) or the first (C-A-G-M20) or second half (M-I-D-R50) of that space. No clear benefit of one grouping over the other in ML performance can be identified
With the \HI{} morphometric space and rather simple machine learning tools, one can identify populations of galaxies that are undergoing ram-pressure stripping reasonably well (precision and recall of $\sim80$\%). This may be useful to construct samples for further inspection but the performance is not yet good enough to reliably infer fractions of galaxies undergoing ram-pressure stripping throughout a survey without further checks. One of our original goals was to ascertain whether this \HI{} morphometric parameter space is good enough to aid in the identification of RPS galaxies. This appears to be feasible. 
Both larger test samples and possible direct use of the \HI{} maps as input may improve results in the future.

\begin{itemize}
    \item The morphometrics can be derived, even with the marginally resolved sources typical for WALLABY (Figure \ref{f:corner:hi}).
    \item They are a reasonably parameter space to identify galaxies that may be undergoing ram pressure (Figure \ref{f:corner:hi}. 
    \item Using an \HI{} contour for GALEX nearby galaxy data is a good segmentation choice (Figure \ref{f:fuv:examples}).
    \item If the whole morphometric space is needed or a subset is viable is not yet clear (Table \ref{t:tree-performance} -- \ref{t:rf}).
    \item A random forest is the best algorithm to classify based on the morphometric space. This has been found on JWST morphology papers as well independently. This saves substantial time/effort compared to Convolutional Neural Networks etc. (Figure \ref{f:matrices} and Table \ref{t:rf}) 
\end{itemize}
For the full WALLABY survey \citep{Koribalski20}, one could apply these to all the solidly detected and partially resolved (e.g. 3-4 beams across) objects to create a parameter space for machine learning to identify populations such as galaxies undergoing ram-pressure stripping or gravitational interactions.

\section{Acknowledgements}

The Australian SKA Pathfinder is part of the Australia Telescope National Facility which is managed by CSIRO. Operation of ASKAP is funded by the Australian Government with support from the National Collaborative Research Infrastructure Strategy. ASKAP uses the resources of the Pawsey Supercomputing Centre. Establishment of ASKAP, the Murchison Radio-astronomy Observatory and the Pawsey Supercomputing Centre are initiatives of the Australian Government, with support from the Government of Western Australia and the Science and Industry Endowment Fund. We acknowledge the Wajarri Yamatji as the traditional owners of the Observatory site.

This research made use of Astropy, a community-developed core Python package for Astronomy \citep{Astropy-Collaboration13,Astropy-Collaboration18}.

AB acknowledges support from the Centre National d’Etudes Spatiales (CNES), France.

SHOH acknowledges a support from the National Research Foundation of Korea (NRF) grant funded by the Korea government (Ministry of Science and ICT: MSIT) (No. NRF-2020R1A2C1008706).

FB acknowledges funding from the European Research Council (ERC) under the European Union’s Horizon 2020 research and innovation programme (grant agreement No.726384/Empire)

BWH is supported by an Enhanced Mini-Grant (EMG). The material is based upon work supported by NASA Kentucky under NASA award No: 80NSSC20M0047.

BWH would like to thank Robin Allen and the teachers at Engelhart elementary for their hard work on distance learning during the COVID-19 lockdown and the game Minecraft for allowing for enough additional time to write this manuscript.

\section{Data Availability}

Electronic versions of the \HI{} and UV morphometric catalogs and the segmentation map are included as supplemental information for this manuscript. 

The full 30 square degree spectral line cubes can be
accessed on the CSIRO ASKAP Science Data Archive\footnote{https://data.csiro.au/domain/casdaObservation} \citep[CASDA,][]{Chapman15, Huynh20} 
with the DOI: \url{https://doi.org/10.25919/5f7bde37c20b5} for this data-set.

%\bibliographystyle{mnras}
%\bibliography{Bibliography} % if your bibtex file is called example.bib
%%%%%%%%%%%%%%%%%%%%%%%%%%%%%%%%%%%%%%%%%%%%%%%%%%

%%%%%%%%%%%%%%%%% APPENDICES %%%%%%%%%%%%%%%%%%%%%

% \appendix

% \section{Catalogs}

% Electronic versions of the \HI{} and UV morphometric catalogs and the segmentation map are included in the online version of this manuscript. 

% Don't change these lines
\bsp	% typesetting comment
\label{lastpage}
\end{document}